\documentclass[twocolumn,showpacs,preprintnumbers,notitlepage,amsmath,amssymb,prb]{revtex4-1}
\usepackage{graphicx}
\usepackage[dvipsnames]{xcolor}
\usepackage{amsmath,amsthm,amssymb}
\usepackage{braket}
\usepackage{multirow}

\usepackage[colorlinks=true,allcolors=red,citecolor=blue]{hyperref}

\begin{document}

\title{Interlayer Coupling Induced Quasiparticles}
\author{Christopher Lane}
\email{laneca@lanl.gov}
\affiliation{Theoretical Division, Los Alamos National Laboratory, Los Alamos, New Mexico 87545, USA}
\affiliation{Center for Integrated Nanotechnologies (CINT-LANL), Los Alamos Laboratories, Los Alamos, New Mexico 87545, USA} 

\begin{abstract} % abstract
We present an exact treatment of layered many-body electronic systems in the presence of interlayer coupling within the Schwinger functional derivative approach on the Keldysh contour. Our transparent approach allows us to clarify the definition of interlayer coupling by showing the independent roles hybridization and interactions play in generating new electronic and magnetic excitations. We find interlayer coupling to induce a variety of plasmons, magnons, and excitons, residing within the a layer, traversing between layers, or propagating along the interface. Moreover, we predict interfacial excitations, including an electron-hole pairing pathway, facilitated by previously ignored layer nonconserving interactions. Finally, we briefly explore the consequence of interlayer coupling on a bilayer square lattice system.
\end{abstract}

\maketitle

%----------------%
\section{Introduction}

Atomically thin 2D materials have proven to be one of the most exciting platforms exhibiting an extensive range of novel electronic\cite{kim2015observation}, excitonic\cite{pollmann2015resonant}, valley\cite{rivera2016valley}, topological\cite{bansil2016colloquium}, and correlated physics.\cite{costanzo2016gate} By combining the 2D building blocks into vertical or lateral heterostructures one may rationally engineer complex multilayer systems and artificial solids with new emergent properties giving way to direct applications in quantum information technologies\cite{aharonovich2016solid,vitale2018valleytronics}, spin optoelectronic devices\cite{cheng2019recent,vitale2018valleytronics}, and energy storage\cite{shi2017recent,lane2019understanding}. To design and manipulate these novel layered materials a detailed theoretical description of the charge, spin, orbital, and layer degrees of freedom is crucial. However, despite vigorous experimental efforts, the development of theoretical techniques going beyond the Hohenberg-Kohn-Sham density functional theory to capture interactions in a layer dependent manner has been slow. In particular, one of the most important first-principles many-body methods used in theoretical spectroscopy for describing excitations involved in radiation-matter interaction is the so-called GW and Bethe-Salpeter equation (BSE), is still awaiting an extension to layer dependent interactions.

The strength of interlayer coupling plays a key role in shaping the emergent properties of heterostructures composed of 2D thin films. For example, when layers are weakly coupled, the absorption profiles of the individual layers is modified,\cite{yuan2018photocarrier} along with the  Raman vibrational modes.\cite{ding2018understanding} In the intermediate regime, the generation of new excitons (interlayer and moir\'e) \cite{ross2017interlayer,zhang2018moire,hennighausen2019oxygen} is facilitated along with the stabilization of superconducting phases.\cite{cao2018unconventional} Lastly, in the limit of strong interlayer coupling, robust charge redistribution is induced\cite{hennighausen2019evidence,yoo2019atomic,lu2019superconductors,cao2018correlated} and the electronic structure of the heterostructure differs considerably from its constituent layers.\cite{yoo2019atomic,hamer2019indirect,vargas2017tunable} 

The influence of interlayer coupling extends beyond atomically-thin 2D materials, playing a significant part in layered transition-metal oxides. In the Ruddlesden-Popper perovskite crystal structure, which includes the cuprate and iridate material families, the two-dimensional perovskite planes are interwoven with layers of alkaline earth, or rare earth metals, and are believed to behave electronically independent. \cite{rao1998transition,kastner1998magnetic} However, a diversity of optimal transition temperatures is observed in the high-temperature cuprate superconductors which appears to be driven at least in part by the choice of rock-salt layer separating the CuO$_2$ planes. For example, the highest Tc obtained in La$_{2-x}$Sr$_x$CuO$_4$ is 40k, whereas in the single layer Hg cuprate, HgBa$_2$CuO$_4$ the optimal Tc is almost 100K,\cite{raghu2012optimal} suggesting that the interlayer interactions between the CuO$_2$ planes and the HgO$_2$ charge-reservoir help to enhance Tc. 

Previous studies using many-body perturbation theory on layered electron gas systems\cite{hawrylak1987effective,hawrylak1988many,white1989many,mahan1990dielectric,white1991many,alatalo1993collective} found the electron effective mass and quasiparticle life time  gave qualitatively different results compared to isolated two- and three-dimensional systems. These models consist of a many-electron Hamiltonian with a Coulomb interaction only, where the electron-electron interaction were restricted to be with in a single layer. Therefore, there is currently no theory that addressees the many-body dynamics arising from the full spin and layer dependent interactions.

In this paper we present an exact treatment of layered many-body electronic systems within the Schwinger functional derivative technique on the Keldysh contour. An advantage to working within the Schwinger Green's function approach is to enable direct access to spectroscopic relevant quantities and therefore, enabling direct comparison and interpretation of experimental spectra. Our results clarify the definition of interlayer coupling by showing the independent roles hybridization and interactions play in generating new electronic and magnetic excitations. By examining the charge and magnetic response functions, along with the Bethe-Salpeter equation for the two-particle Green's function, we predict interfacial plasmons, magnons, and excitons facilitated by layer nonconserving interactions. We briefly explore the consequence of interlayer coupling on a bilayer square lattice system.

\section{Theory}\label{Sec:theory}
\subsection{Hamiltonian and Basic Notations}
The Hamiltonian for a layered system with spin and layer dependent interactions is given by 
\begin{widetext}
\begin{align}
\hat{\mathcal{H}}&=\sum_{\substack{\alpha l\\\beta l^{\prime}}} \int d^{3}r 
\hat{\psi}^{\dagger}_{\alpha l}(r)
h^{0}_{\alpha l,\beta l^{\prime}}(r) 
\hat{\psi}_{\beta l^{\prime}}(r)
+
\frac{1}{2}\sum_{\substack{\alpha\beta\gamma\delta\\ i j k l }}\int \int d^{3}rd^{3}r^{\prime} 
\hat{\psi}^{\dagger}_{\alpha i}(r) \hat{\psi}^{\dagger}_{\beta j}(r^\prime) 
v_{\delta\gamma;\alpha\beta}^{lk;ij}(r,r^{\prime})
\hat{\psi}_{\gamma k}(r^\prime) \hat{\psi}_{\delta l}(r)
\end{align}
\end{widetext}
where the Greek  and Latin letters denote the spin  and layer degrees of freedom, respectively. Our interaction index notion follows an {\it in$_{r}$in$_{r^{\prime}}$};{\it out$_{r}$out$_{r^{\prime}}$} scheme inline with the diagrammatic representation. For an $N$ layer system the Hamiltonian of the $l^{th}$ layer is given by $h^{0}_{\alpha l,\beta l}(r)$. If the layers are close enough for the wave functions of adjacent layers to overlap, electrons can hop from one layer to another. The amplitude of hopping from layer $l^{\prime}$ to layer $l$ is $h^{0}_{\alpha l,\beta l^{\prime}}(r)$. Here, $\mathbf{r}$ is defined over $\mathbb{R}^3$ and the field operators acting on a specific layer $l$  can be written as $\hat{\psi}_{l}(\mathbf{r})\equiv\hat{\psi}(\mathbf{r}+\mathbf{R}_{l})$, where $\mathbf{R}_{l}$ is perpendicular to the interface between the layers and is the distance of the $l^{th}$ layer from the origin layer, $\mathbf{R}_{0}=0$.

\begin{figure}[h!]
\includegraphics[width=0.99\columnwidth]{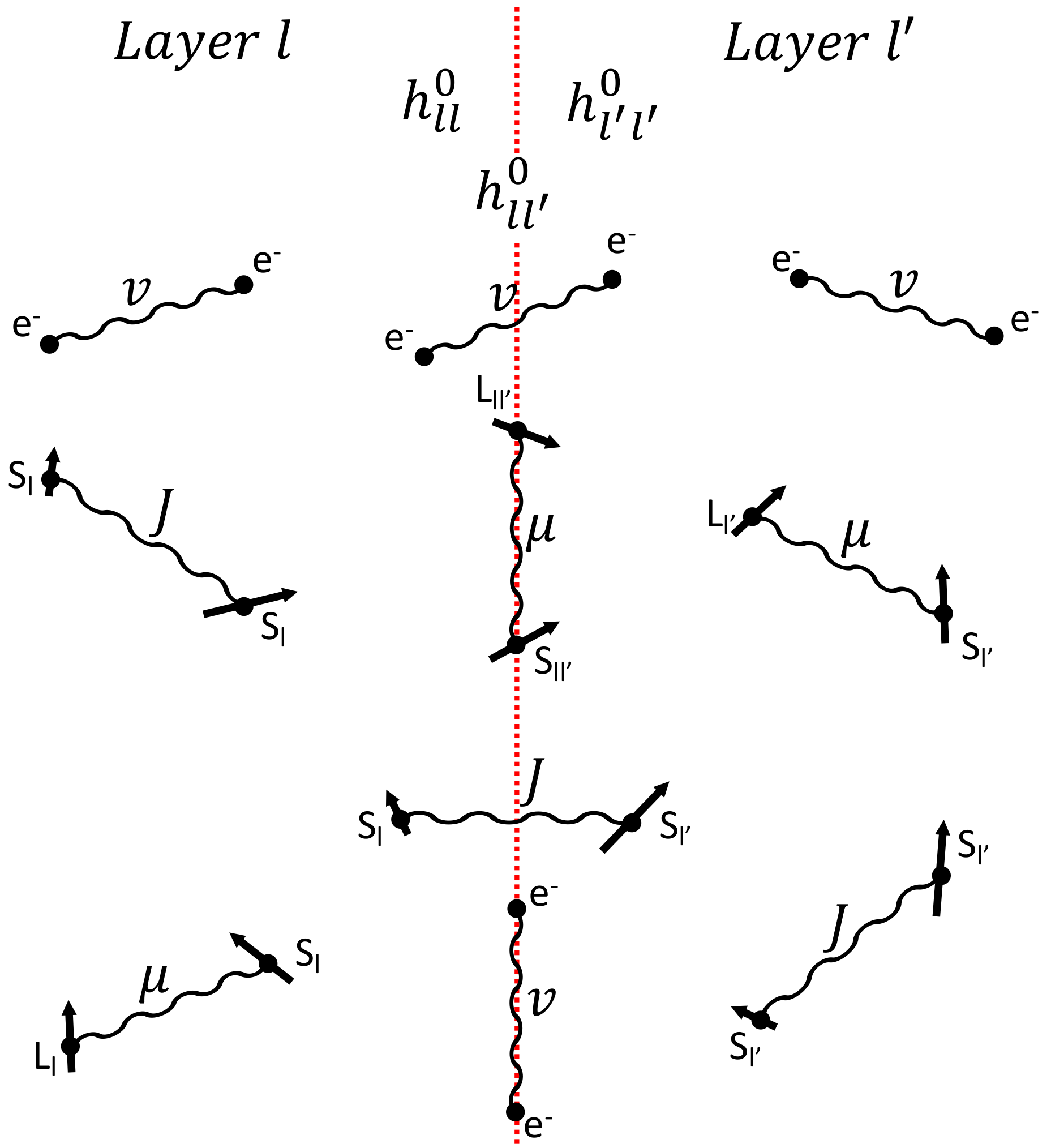}
\caption{(color online) A schematic representation of the Coulomb, spin-spin, and spin-orbit interactions in real space within and between layer $l$ and $l^{\prime}$, along with each layer's internal electronic structure $h^{0}_{\alpha l,\beta l}(r)$ and hybridization  $h^{0}_{\alpha l,\beta l^{\prime}}(r)$. The red dashed line denotes the boundary between layer $l$ and $l^{\prime}$.} 
\label{fig:HAM}
\end{figure}

The generalized two-particle interaction takes both the spin and layer configuration into account and can be broken down into three contributions: 
\[ v_{\delta\gamma;\alpha\beta}^{lk;ij}(r,r^{\prime})=
   \begin{cases} 
      \sigma^{0}_{\alpha\delta}v^{lk;ij}(r,r^{\prime})\sigma^{0}_{\beta\gamma} \\
      \sigma^{I}_{\alpha\delta}J^{lk;ij}_{IJ}(r,r^{\prime})\sigma^{J}_{\beta\gamma}\\
      \sigma^{I}_{\alpha\delta}\mu^{lk;ij}_{I}(r,r^{\prime})\sigma^{0}_{\beta\gamma}
   \end{cases}.
\]
The first is the usual Coulomb interaction, the second a spin-spin interaction, and the third a spin-orbit interaction. The layer degrees of freedom can be classified based on their vertex. In analogy to spin, the vertex maybe layer number conserving or nonconserving, giving way to interactions originating within the same layer, between layers, or at the interface, as schematically illustrated in Fig.~\ref{fig:HAM}. Table~\ref{table:interactionClass} gives the various classes of interactions. The existence of interfacial interactions is a direct consequence of the boundary between the various layers, where the boundary acts as an impurity by flipping the conserved layer quantum numbers. In previous works on layered electron gases the role of these interfacial interactions has been ignored.\cite{hawrylak1987effective,hawrylak1988many,white1989many,mahan1990dielectric,white1991many,alatalo1993collective}  

Later on it will be helpful to expand the spin degrees of freedom in the Pauli and identity matrices as
\begin{align}\label{eq:spinchangeofbasis}
v_{\delta\gamma;\alpha\beta}^{lk;ij}(r,r^{\prime})=\sigma^{I}_{\alpha\delta}v_{IJ}^{lk;ij}(r,r^{\prime})\sigma^{J}_{\beta\gamma},
\end{align}
where $I,J\in \{0,x,y,z\}$.
In strongly spin-orbit coupled systems, e.g., heavy fermion systems, the two-particle interaction can be modified to consider $J\cdot J$ coupling rather than the Russell-Saunders $L\cdot S$ coupling.\cite{leighton1959,freeman1967} Additionally, the layer degrees of freedom maybe expanded into the identity plus the generators of SU($N$), where $N$ is the number of layers. For example, in a bilayer system the layer indices can be reorganized using the Pauli matrices while for a trilayer system the Gell-Mann matrices are used.

\begin{table}[h]
\centering
{\renewcommand{\arraystretch}{1.8}
\begin{tabular}{c|c|c|c|}
& Coulomb & Spin & Spin-Orbit \\
\hline
Intralayer & $v_{00}^{ll;ll}$ & $v_{IJ}^{ll;ll}$ & $v_{0J}^{ll;ll}$\\
Interlayer & $v_{00}^{lk;lk}$ & $v_{IJ}^{lk;lk}$ & $v_{0J}^{lk;lk}$ \\
\hline
Interfacial & $v_{00}^{lk;l^\prime k^\prime}$ & $v_{IJ}^{lk;l^\prime k^\prime}$ & $v_{0J}^{lk;l^\prime k^\prime}$ \\
\end{tabular}
}
\caption{Classification of the various electron-electron interactions by layer and charge-spin degrees of freedom. The first two rows present layer-number conserving interactions, while the last row is layer nonconserving.}\label{table:interactionClass}
\end{table}

To keep the results and the discussion general we define all operators in the {\it contour} Heisenberg picture,
\begin{align}
\mathcal{O}(z)=U(z_0,z)\mathcal{O}U(z,z_0),
\end{align}
with the time arguments, $z$ and $z_0$, running along the Keldysh contour $(z\in \mathcal{C})$, where $z_{0}$ is an arbitrary initial time and the time evolution operator, $U(z_0,z)$, evolves an operator  $\mathcal{O}$ from   $z_{0}$ to $z$ along the contour. In this picture the operators are explicitly time dependent whereas the wave functions are not. This allows us to introduce the time ordering on the contour and Wick's theorem, connecting our results to many-body perturbation theory.\cite{stefanucci2013}

In order to treat the electronic many-body dynamics at finite temperature, we define the time-dependent ensemble average of operator $\mathcal{O}(z)$ as 
\begin{align}
\braket{\mathcal{O}(z)}=\frac{\text{Tr}\left\{  \mathcal{T} \exp{\left[-i\int_{\mathcal{C}}  d\bar{z} H(\bar{z}) \right]    }    \mathcal{O}(z)  \right\}    }{\text{Tr}\left\{  \mathcal{T} \exp{\left[-i\int_{\mathcal{C}}  d\bar{z} H(\bar{z}) \right]    }    \right\}   },
\end{align}
where $\braket{\mathcal{O}(z)}$ is the overlap between the initial state in thermodynamical equilibrium (for temperature $\beta$) at $z_0$ with the time evolved state at $z$.

\footnotetext[999]{For an excellent historical overview of the Schwinger Green's function method and Schwinger's personal retrospective on the influence of Green's functions on his work, see Refs.~\onlinecite{schweber2005sources} and \onlinecite{schwinger1993greening}.  }

To obtain the exact expression for the self-energy, the vertex, and various other quantities we will use the Schwinger functional derivative approach.\cite{schwinger1951greenI,schwinger1951greenII,Note999} To do so, we couple our Hamiltonian to a time dependent auxiliary electromagnetic field that probes the charge, spin, and layer degrees of freedom. The electric and magnetic fields are given in a compact form by 
\begin{align}\label{eq:extfield}
\hat{\pi}(z_{1})=\int d^2r \pi_{l l^{\prime}}^{I}(1)  \hat{\psi}^{\dagger}_{\alpha l}(1)\sigma_{\alpha\beta}^{I} \hat{\psi}_{\beta l^{\prime}}(1).
\end{align}
Now if we wish to find the infinitesimal change in a generic, contour-ordered product of operators $\Pi_{i}\mathcal{O}_{i}(z_{i})$ with respect to field $\pi_{l l^{\prime}}^{I}(1)$, we arrive at the following identity,
\begin{align}\label{eq:funcderivindentity}
i\frac{\delta}{\delta \pi_{l l^{\prime}}^{I}(1)}&\braket{\mathcal{T}\left\{  \Pi_{i}\mathcal{O}_{i}(z_{i})  \right\}   }=\nonumber\\
&\braket{\mathcal{T}\left\{  \Pi_{i}\mathcal{O}_{i}(z_{i})      \hat{\psi}^{\dagger}_{\alpha l}(1)\sigma_{\alpha\beta}^{I} \hat{\psi}_{\beta l^{\prime}}(1)  \right\}   }\nonumber\\
-&\braket{\mathcal{T}\left\{  \Pi_{i}\mathcal{O}_{i}(z_{i})  \right\}   }\braket{\mathcal{T}\left\{     \hat{\psi}^{\dagger}_{\alpha l}(1)\sigma_{\alpha\beta}^{I} \hat{\psi}_{\beta l^{\prime}}(1)   \right\}   }
\end{align}
where $\mathcal{T}$ is the contour-ordering operator. In general this identity is valid for same time and mixed operators, including electronic and bosonic; for more details see Ref. \onlinecite{stefanucci2013}.

\subsection{Generalized Hedin's Equations for a multi-layered spin dependent system}

The derivation closely follows Hedin's original work\cite{hedin1965}, along with other more recent generalizations\cite{marini2018,aryasetiawan1999,aryasetiawan2008}, using Schwinger's functional derivative technique.
Since the fermionic field operator satisfies the Heisenberg equation of motion
\begin{align}\label{eq:psieom}
\frac{d}{dz_{1}}\hat{\psi}_{\alpha n}(1)=i\left[  \mathcal{H}, \hat{\psi}_{\alpha n}(1) \right],
\end{align}
we can straightforwardly derive the equation of motion of the Green's function,
\begin{widetext}
\begin{align}\label{eq:G-EOM}
\left(  i\frac{d}{dz_{1}} \delta_{l^{\prime}n}\delta_{\alpha\beta}-h^{0}_{\alpha n,\beta l^{\prime}}(1)   \right)G_{\beta l^{\prime},\sigma m}(1,2)
=
\delta(1,2)\delta_{\alpha\sigma}\delta_{nm}
-iv^{lk;in}_{\delta\gamma;\xi\alpha}(3,1)G^{(2)}_{\gamma k,\delta l,\xi i,\sigma m}(1,3,3^+,2),
\end{align}
\end{widetext}
where the single- and two-particle Green's functions are given by 
\begin{align}
G_{\beta l^{\prime},\sigma m}(1,2)&=\frac{1}{i}\braket{         \hat{\psi}_{\beta l^{\prime}}(1)       \hat{\psi}^{\dagger}_{\sigma m}(2)   },\\
G^{(2)}_{\gamma k,\delta l, \xi i,\sigma m}(1,3,3^+,2)&=\frac{1}{i^2}  \braket{  \hat{\psi}_{\gamma k}(1) \hat{\psi}_{\delta l}(3)     \hat{\psi}^{\dagger}_{\xi i}(3^+)    \hat{\psi}^{\dagger}_{\sigma m}(2)   },
\end{align}
where $(^+)$ in $ \hat{\psi}^{\dagger}_{\eta j}(3^+)$ denotes this operator should be placed infinitesimally after $\hat{\psi}_{\gamma k}(3) $ when the time ordering operator is applied. The electron creation and annihilation operators were also taken to obey the canonical anticommutation relations on the contour
\begin{align}
\left\{  \hat{\psi}_{\alpha l}(1) , \hat{\psi}^{\dagger}_{\beta l^{\prime}}(2)  \right\}&=\delta_{\alpha\beta}\delta_{l l^{\prime}}\delta(1-2),\\
\left\{  \hat{\psi}_{\alpha l}(1) , \hat{\psi}_{\beta l^{\prime}}(2)  \right\} &= 0 = \left\{  \hat{\psi}^{\dagger}_{\alpha l}(1) , \hat{\psi}^{\dagger}_{\beta l^{\prime}}(2)  \right\},
\end{align}
where we have introduced the short hand $\hat{\psi}^{\dagger}_{\beta l^{\prime}}(2) \equiv \hat{\psi}^{\dagger}_{\beta l^{\prime}}(\mathbf{x}_{2},z_{2}  )$. For convenience we  use the convention where a repeated index or variable implies a summation or integration, provided the repeated indices are on the same side of the equation.

Originally, Hedin just considered an external electric field which was used to relate the two-particle Green's function to the functional derivative of the single particle Green's function with respect to the probing electric field. Here, we have coupled our Hamiltonian to a layer dependent electromagnetic field allowing us to capture the intertwined charge, spin, and layer excitations of the system. Therefore, by Eq.~(\ref{eq:funcderivindentity}), the two-particle Green's function can be written as
\begin{align}\label{eq:2partdecomp}
&G^{(2)}_{\gamma k,\delta l, \xi i,\sigma m}(1,3,3^+,2)\sigma_{\xi\delta}^{I}=\nonumber\\
&G_{\gamma k,\sigma m}(1,2)   G_{\delta l,\xi i}(3,3^+) \sigma^{I}_{\xi \delta} -  \frac{\delta G_{\gamma k,\sigma m}(1,2)}{\delta \pi^{I}_{il}(3)}.
\end{align}
From this relation we recover the mass operator,
\begin{align}
\mathcal{M}_{\alpha n,\nu t}(1,3)&G_{\nu t,\sigma m}(3,2)=\nonumber\\
&-i\sigma_{\xi\delta}^{I}v^{lk;in}_{IJ}(3,1)\sigma_{\alpha\gamma}^{J}G^{(2)}_{\gamma k,\delta l, \xi i,\sigma m}(1,3,3^+,2)\nonumber
\end{align}
\begin{align}
=V^{J}_{H~k;n}(1)\sigma^{J}_{\alpha\gamma}&G_{\gamma k,\sigma m}(1,2)\nonumber\\
&+\Sigma_{\alpha n,\nu t}(1,5)G_{\nu t,\sigma m}(5,2).
\end{align}

Two contributions can be readily identified, the generalized Hartree potential 
\begin{align}
V^{J}_{H~k;n}(1)=\rho^{I}_{il}(3) v^{lk;in}_{IJ}(3,1) ,
\end{align}
and the exact self-energy
\begin{align}
\Sigma_{\alpha n,\nu t}(1,5)=-iv^{lk;in}_{IJ}(3,1)\sigma_{\alpha\gamma}^{J}G_{\gamma k,\mu s}(1,4)\frac{\delta G^{-1}_{\mu s,\nu t}(4,5)}{\delta \pi^{I}_{il}(3)},
\end{align}
where we have used the identity,
\begin{align}\label{eq:identityGG-1G}
\frac{\delta \left(G^{-1}G\right)}{\delta \pi}=0 \implies \frac{\delta G}{\delta \pi}=-G\frac{\delta G^{-1}}{\delta \pi}G,
\end{align}
to pull out a factor of $G$ along with the definition of the charge and spin density,
\begin{align}
\rho^{I}_{il}(3)=-iG_{\delta l,\xi i}(3,3^+) \sigma^{I}_{\xi \delta}.
\end{align}

To uncover the richness of the self-energy, we expand\footnote{By the chain rule \begin{align}
\frac{\delta G^{-1}_{\mu s,\nu t}(4,5)}{\delta \pi^{I}_{il}(3)}
=
\frac{\delta G^{-1}_{\mu s,\nu t}(4,5)}{\delta \Phi^{L}_{ab}(6)}\frac{\delta \Phi^{L}_{ba}(6)}{\delta \pi^{I}_{il}(3)},
\end{align}
where the first and second terms contribute to the vertex and screened interaction, respectively.} the functional derivative in the self-energy in terms of the total field
\begin{align}
\Phi^{J}_{nk}(1)=\pi^{J}_{nk}(1)+V^{J}_{H~k;n}(1).
\end{align}
We find, just as Hedin, that the self-energy is made up of three interwoven components: the single-particle Green's function, the screened interaction, and the vertex, 
\begin{align}
\Sigma_{\alpha n,\nu t}(1,5)=i\sigma_{\alpha\gamma}^{J}G_{\gamma k,\mu s}(1,4)  \Lambda^{L~ab}_{\mu s,\nu t}(4,5;6)W^{LJ}_{ak;bn}(6,1).
\end{align}
The vertex is 
\begin{align}
\Lambda^{L~ab}_{\mu s,\nu t}(4,5;6)=-\frac{\delta G^{-1}_{\mu s,\nu t}(4,5)}{\delta \Phi^{L}_{ab}(6)}
\end{align}
and the screened interaction is
\begin{subequations}
\begin{align}
W^{LJ}_{ak;bn}(6,1)&= \frac{\delta \Phi^{L}_{ba}(6)}{\delta \pi^{I}_{il}(3)} v^{lk;in}_{IJ}(3,1),\\
&=\varepsilon^{-1~LI}_{al;bi}(6,3)v^{lk;in}_{IJ}(3,1).
\end{align}
\end{subequations}

To find the self-consistent equations governing $W$ and $\Lambda$, we use the equation of motion of $G$ and employ the chain rule,
\begin{align}
W&^{LJ}_{ak;bn}(6,1)=\varepsilon^{-1~LI}_{al;bi}(6,3)v^{lk;in}_{IJ}(3,1)\\[0.6em]
=&\left(  \delta(6,3)\delta_{LI}\delta_{bi}\delta_{al} +
\frac{\delta V^{L}_{H~a;b}(6)}{\delta \rho^{M}_{cd}(7)}\frac{\rho^{M}_{dc}(7)}{\delta \Phi^{N}_{fg}(8)}\frac{\delta \Phi^{N}_{gf}(8)}{\delta \pi^{I}_{il}(3)}
\right)v^{lk;in}_{IJ}(3,1)\nonumber\\[0.6em]
=& v^{ak;bn}_{LJ}(6,1) +
v_{LM}^{ad;bc}(6,7)\chi_{0~cf;dg}^{MN}(7,8)
W^{NJ}_{fk;gn}(8,1),\nonumber\\ \nonumber
\end{align}
where we used the indistinguishability of particles, $v^{da;cb}_{ML}(7,6)=v^{ad;bc}_{LM}(6,7)$.
The vertex expends as 
\begin{align}\label{eq:vertex}
\Lambda&^{L~ab}_{\alpha n , \eta y}(1,4;6)=\delta(1,6)\delta(1,4)\sigma^{L}_{\alpha\eta }\delta_{an}\delta_{by}
+\frac{\delta \Sigma_{\alpha n , \eta y}(1,4)}{\delta \Phi^{L}_{ab}(6)}\nonumber\\
&=\delta(1,6)\delta(1,4)\sigma^{L}_{\alpha\eta }\delta_{an}\delta_{by}\nonumber\\
&+\frac{\delta \Sigma_{\alpha n , \eta y}(1,4)}{\delta G_{\mu s , \nu t}(9,10)}G_{\nu t , \epsilon g}(9,11)\Lambda^{L~ab}_{\epsilon g , \delta f}(11,12;6)G_{\delta f , \mu s }(12,10).
\end{align}
Additionally, we define the polarization as 
\begin{align}\label{eq:polarization}
\chi&_{0~cf;dg}^{MN}(7,8)=\frac{\rho^{M}_{dc}(7)}{\delta \Phi^{N}_{fg}(8)}\\
&=-iG_{\delta c , \mu s}(7,9)\Lambda^{N~fg}_{\mu s,\nu t}(9,10;8) G_{\nu t , \xi d}(10,7^+)\sigma^{M}_{\xi \delta}.\nonumber
\end{align}

\begin{figure}[h!]
\includegraphics[width=\columnwidth]{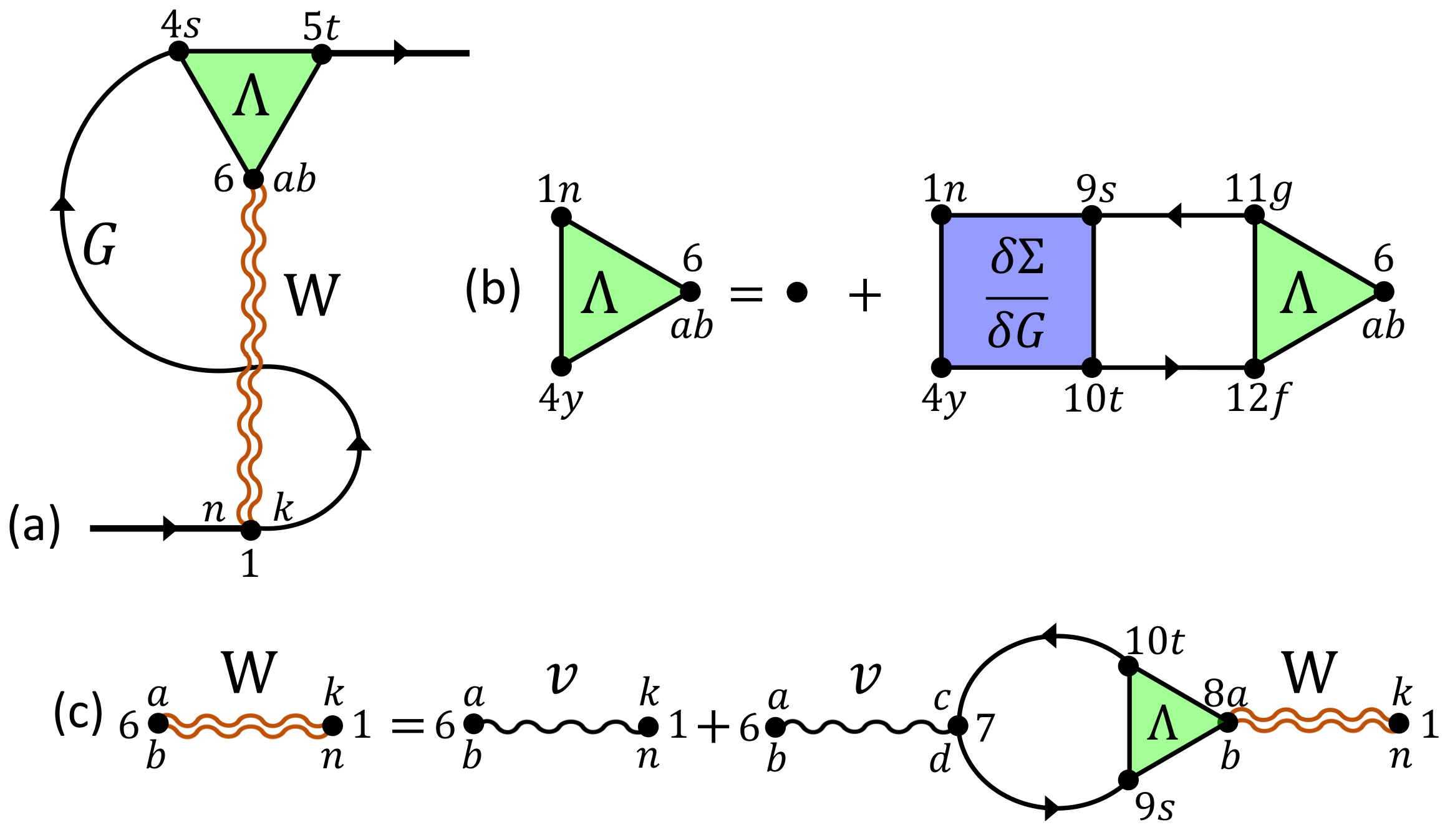}
\caption{(color online) Diagrammatic representation of the self-energy (a), the vertex (b), and the screened interaction (c) functions.  The spin indices have been suppressed for clarity.}
\label{fig:diagrams}
\end{figure}

The complete set of self-consistent layer and spin dependent Hedin's equations relating the electronic self-energy $\Sigma$ to the Green's function $G$ and the screened interaction $W$, using the vertex $\Lambda$ and polarization function $\chi_0$ are:
\begin{widetext}
\begin{subequations}
\begin{align}\label{eq:fullsetin}
\Sigma&_{\alpha n,\nu t}(1,5)=i\sigma_{\alpha\gamma}^{J}G_{\gamma k,\mu s}(1,4)  \Lambda^{L~ab}_{\mu s,\nu t}(4,5;6)W^{LJ}_{ak;bn}(6,1),\\
\nonumber\\
W&^{LJ}_{ak;bn}(6,1)= v^{ak;bn}_{LJ}(6,1) + v_{LM}^{ad;bc}(6,7)\chi_{0~cf;dg}^{MN}(7,8)W^{NJ}_{fk;gn}(8,1),\\
\nonumber\\
\chi&_{0~cf;dg}^{MN}(7,8)=-iG_{\delta c , \mu s}(7,9)\Lambda^{N~fg}_{\mu s,\nu t}(9,10;8) G_{\nu t , \xi d}(10,7^+)\sigma^{M}_{\xi \delta},\\
\nonumber\\
\Lambda&^{L~ab}_{\alpha n , \eta y}(1,4;6)=\delta(1,6)\delta(1,4)\sigma^{L}_{\alpha\eta }\delta_{an}\delta_{by}+\frac{\delta \Sigma_{\alpha n , \eta y}(1,4)}{\delta G_{\mu s , \nu t}(9,10)}G_{\nu t , \epsilon g}(9,11)\Lambda^{L~ab}_{\epsilon g , \delta f}(11,12;6)G_{\delta f , \mu s }(12,10).\label{eq:fullsetout}
\end{align}
\end{subequations}
\end{widetext}
To close the set of equations, Dyson's equation 
\begin{align}
G&_{\alpha n,\beta m}(1,2)=\\
&G_{0~\alpha n,\beta m}(1,2)+G_{0~\alpha n,\eta s}(1,3) \Sigma_{\eta s,\delta l}(3,4)G_{\delta l,\beta m}(4,2),\nonumber
\end{align}
links the fully interacting system to the bare noninteracting propagator,
\begin{align}
G^{-1}&_{0~\alpha n,\eta y}(1,4)=\\
&\left( i\frac{d}{dz_{1}} \delta_{\alpha\eta}\delta_{yn} - h^{0}_{\alpha n , \eta y}(1)-\Phi^{N}_{ny}(1)\sigma^{N}_{ \alpha \eta} \right)\delta(1,4).
\end{align}
A diagrammatic representation of these equations is shown in Fig.~\ref{fig:diagrams}.

\begin{figure}[h!]
\includegraphics[width=\columnwidth]{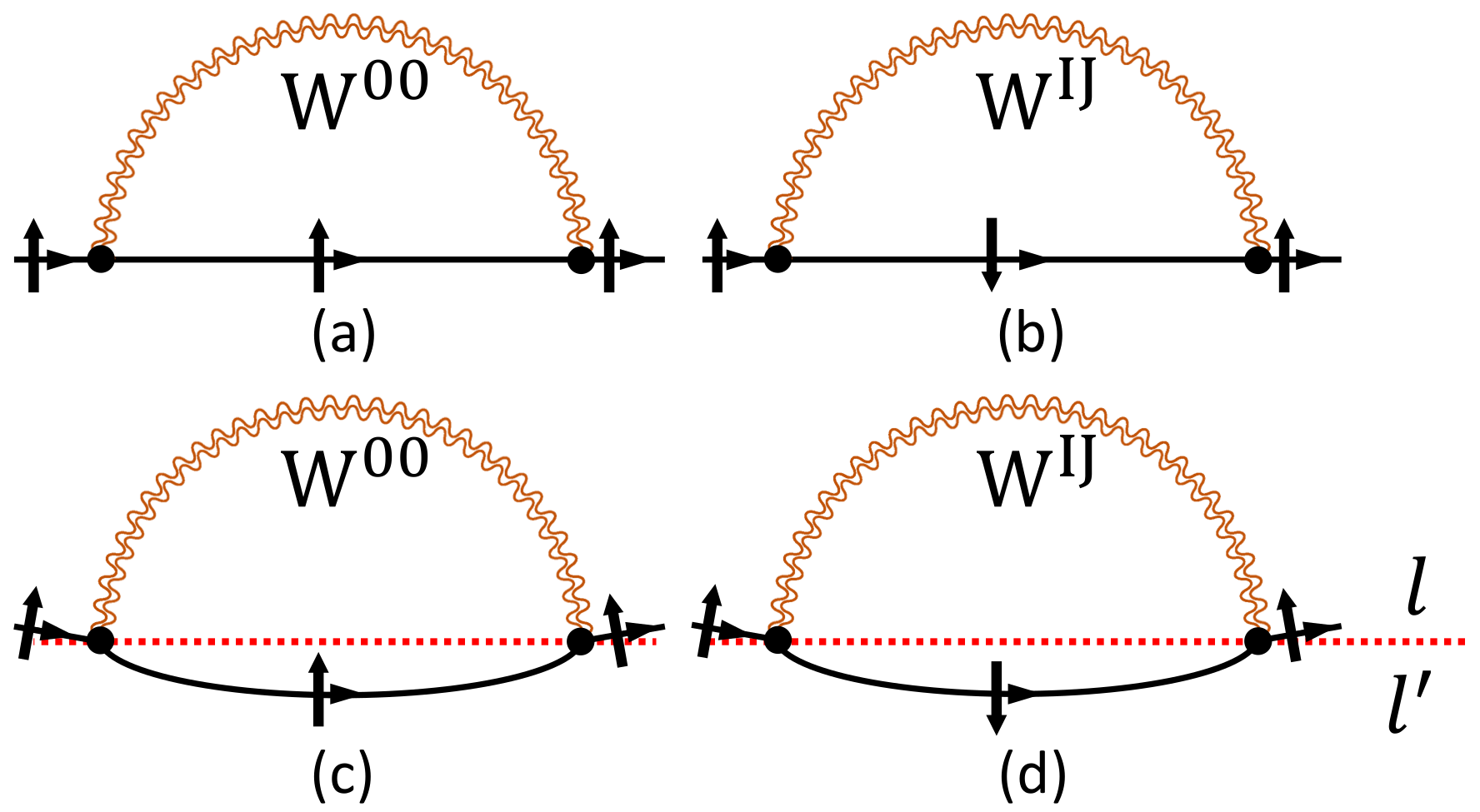}
\caption{(color online) Diagrammatic representation of the self-energy in the $GW$ approximation for layer conserving (a),(b) and layer nonconserving (c),(d) interactions. The right hand (left hand) diagrams show an electron exchanging energy and momentum with plasmons (magnons) represented by $W^{00}$ ($W^{IJ}$). The boundary between layers $l$ and $l^{\prime}$ is indicated by the red dashed line.}
\label{fig:GWtypes}
\end{figure}

Before moving forward, let use interpret the meaning of the resulting self-energy and screened interaction. For simplicity and clarity, we will take
\begin{align}\label{eq:bareVertex}
\Lambda&^{L~ab}_{\alpha n , \eta y}(1,4;6)=\delta(1,6)\delta(1,4)\sigma^{L}_{\alpha\eta }\delta_{an}\delta_{by},
\end{align}
which yields the commonly employed GW approximation\cite{onida2002electronic}, where
\begin{align}\label{eq:GWapprox}
\Sigma&_{\alpha n,\nu t}(1,5)=i\sigma_{\alpha\gamma}^{J}G_{\gamma k,\mu a}(1,5)\sigma^{L}_{\mu \nu}W^{LJ}_{ak;tn}(5,1),
\end{align}
and
\begin{align}
\chi&_{0~dg;cf}^{MN}(7,8)=-iG_{\delta c, \mu f}(7,8)\sigma^{N}_{\mu\nu}G_{\nu g , \xi d}(8,7^+)\sigma^{M}_{\xi \delta}.
\end{align}
If the hybridization between layer $l$ and $l^{\prime}$ is assumed to be small, as is the case for vertical heterostructures composed of 2D transition metal dichalcogenides, \cite{miro2014atlas,manzeli20172d} the self-energy can be partitioned into two classes involving layer conserving and layer nonconserving interactions. We illustrate the physical meaning of each case in the following. 

{\it Layer Conserving}: If a particle, $G_{\uparrow l,\uparrow l}$, of up-spin and layer $l$ enters the self energy $\Sigma_{\uparrow l,\uparrow l}$, the particle exchanges energy and momentum with plasmons, $W^{00}$. If the same particle on layer $l$ has its spin flipped upon entering the self-energy by a spin operator $\sigma^{I}_{\uparrow\downarrow}$, a magnon given by $W^{IJ}$ is emitted. Then upon exiting the self-energy the magnon is reabsorbed, thereby flipping the spin by $\sigma^{J}_{\downarrow\uparrow}$, and recovering its original spin state.  We call these {\it intra}-layer plasmons (magnons). This process is illustrated in Figs.~\ref{fig:GWtypes}(a) and \ref{fig:GWtypes}(b). 

{\it Layer Nonconserving}: If a particle, $G_{\uparrow l,\uparrow l}$, of up-spin and layer $l$ enters the self energy $\Sigma_{\uparrow l,\uparrow l}$, the screened interaction, $W$, can `flip' the layer on which the particle is propagating, as seen in Figs.~\ref{fig:GWtypes} (c) and \ref{fig:GWtypes}(d). As the particle changes layer, it can also emit a plasmon (magnon). On exiting the self energy, the particle is sent back to its layer of origin and reabsorbs the formally emitted plasmon (magnon). We call these {\it interfacial} plasmons (magnons), since these bosonic excitations run along the interface.

Since $G$ is assumed to be nearly diagonal in layer in the weak hybridization limit, the screened interaction only permits two types of polarization bubbles,
\begin{align}\label{eq:bubbles}
\chi_{0~ll;ll}^{MN}(7,8)~~~\text{and}~~~\chi_{0~kl;lk}^{MN}(7,8).
\end{align}
The first bubble completely resides within a single layer and is only able to connect to layer conserving vertices. In contrast, the second polarization bubble is composed of an electron and hole residing on different layers and can only be stimulated by layer nonconserving vertices. 

\newpage
\onecolumngrid

\begin{figure}[h!]
\includegraphics[width=.99\textwidth]{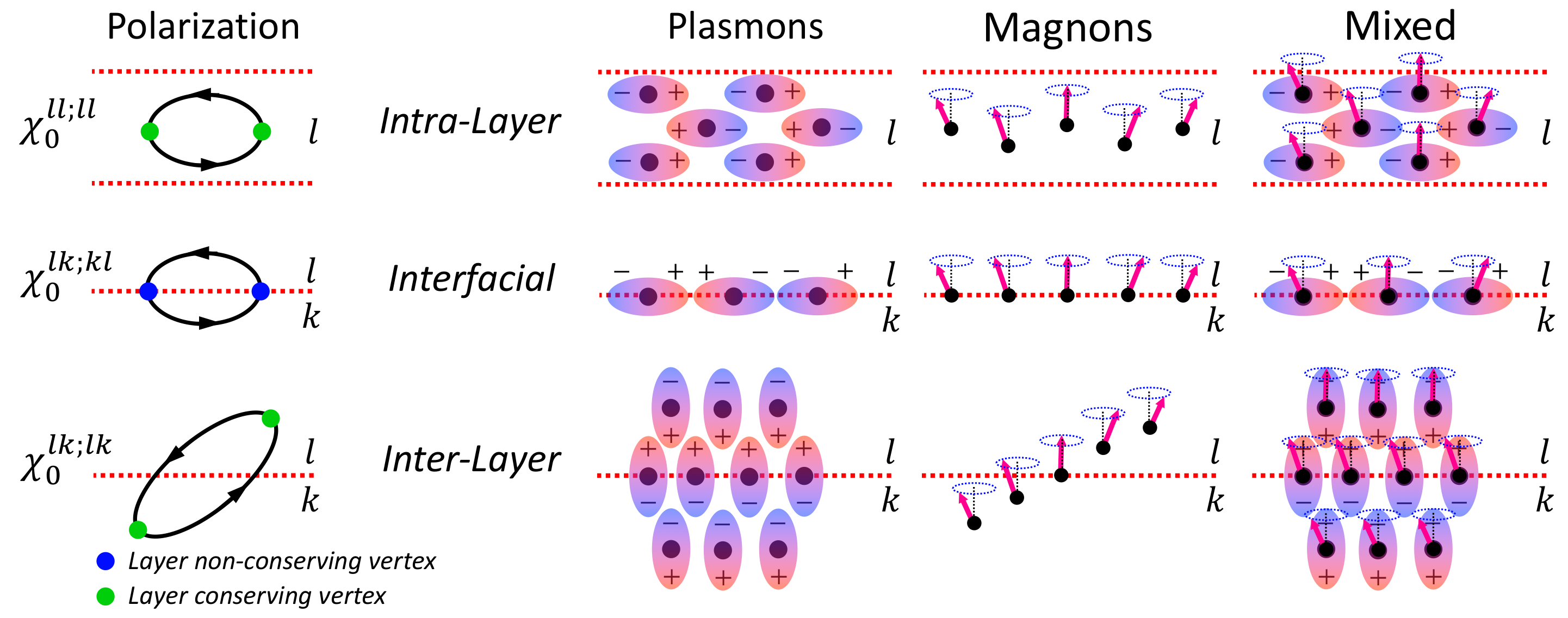}
\caption{(color online) Various polarization bubbles and collective modes induced by interlayer coupling. (Left panel) Polarization bubbles present in the weak and intermediate hybridization limit connecting to layer conserving or layer nonconserving interactions, indicated by the green and blue vertices, respectively. (Right panel) The various bosonic modes predicted as poles of Eq.~(\ref{eq:defchiRStoner}). Intra- and interlayer plasmons (magnons) are facilitated by layer conserving interactions, while the interfacial modes are generated by layer nonconserving interactions. Mixed modes are generated in the presences of spin-orbit coupling or noncollinear magnetic ordering. The boundary between layers is indicated by the red dashed line.}
\label{fig:screeningandbosons}
\end{figure}
\twocolumngrid

For intermediate strength layer hybridizations, such as those in the high-temperature superconducting cuprates\cite{cooper_anisotropy_1994,furness2018accurate,lane2018antiferromagnetic,zhang2020competing} and the perovskite iridates\cite{zhao2018decoupling,lane2020iridate}, an additional interlayer polarization bubble is possible,
\begin{align}
\chi&_{0~lk;lk}^{MN}(7,8).
\end{align}
For this bubble, its vertices reside on different layers, allowing connections to layer conserving interactions only. Recently, this interlayer polarization has been found to contribute to the effective screening of the Ni 3$d$ orbitals from the Nd layer in the newly discovered nickelate superconductor NdNiO$_2$.\cite{nomura2019formation} In the strong hybridization limit all remaining polarization bubble configurations are found and play a role in the full screened interaction. These three bubbles are sketched in Fig.~\ref{fig:screeningandbosons} (left panel).

As elucidated by Perdew {\it et al.},\cite{kurth2000role} the exchange-correlation energy may only account for a small fraction of the total energy of a system, but it includes three key physical ingredients: The exchange energy corrects spurious effects of self-interaction and also maintains the Pauli exclusion principle, while the correlation energy accounts for Coulomb correlation effects in the many-electron environment. However, most importantly, the exchange-correlation energy plays an extremely vital role in the `glue' that binds atoms together to form molecules and solids.
Here, the same principle extents to layered systems. The self-energy in Eq.~\ref{eq:fullsetin} [or  Eq.~(\ref{eq:GWapprox}) in the $GW$ approximation] provides the  exchange and correlation corrections to the bare noninteracting system. Due to the explicit layer dependence, $\Sigma^{ll^\prime}=\Sigma^{ll^\prime}_{x}+\Sigma^{ll^\prime}_{c}$, one finds two types of `glue', one sticking atoms together within the same layer ($l=l^\prime$) and the other adhering the layers together ($l \neq l^\prime$). 

\subsection{Charge and Magnetic Response}
Interlayer coupling has been shown to play a pivotal role in stabilizing various magnetic orders in a layer dependent manner\cite{huang2017layer} and enhancing interlayer-exchange coupling\cite{klein2019enhancement} between 2D atomically thin films. To analyze the instability of the ground state to various ordered phases and investigate the emergent excitations harbored in layered systems, we must observe its response to an infinitesimal time-dependent external probe $\pi^{I}_{ij}(1)$. The response of the system due to an infinitesimal change in the external field is
\begin{align}\label{eq:defchiR}
\chi^{IJ}_{mj;ni}(1,2)=\frac{\delta \rho^{I}_{nm}(1)}{\delta \pi^{J}_{ij}(2)}.
\end{align}
Using the chain rule we arrive at a recursive relationship for density response due to the perturbation,
\begin{align}\label{eq:defchiR}
\chi&^{IJ}_{mj;ni}(1,2)=\frac{\delta \rho^{I}_{nm}(1)}{\delta \Phi^{M}_{ab}(3)}\frac{\delta \Phi^{M}_{ba}(3)}{\delta \pi^{J}_{ij}(2)}\nonumber\\
&= \chi^{IJ}_{0~mj;ni}(1,2)  +\chi^{IM}_{0~ma;nb}(1,3)   v^{at;bs}_{ML}(3,4)  \chi^{LJ}_{tj;si}(4,2).
\end{align}
To dissect the meaning and structure of this response function we start with its tensoral structure. Due to the generalized nature of the density and external electromagnetic field, the response $\chi^{IJ}_{mj;ni}$ not only contains the charge $(I,J=0)$ and spin $(I,J\in\{ x,y,z\})$  responses, but also their mixture $(I=0,J\in\{ x,y,z\})$. Moreover, each of these responses is indexed by layer $(mj;ni)$ in which the electrons and holes reside or transition between. This fine grained, transparent indexing structure gives us a comprehensive picture of the various responses found in interacting layered systems. 

If we expand the recursive relation in Eq.~(\ref{eq:defchiR}) to a few orders in $v$, one can convince themselves that $\chi^{IJ}_{mj;ni}(1,2)$ is composed of all combinations of polarization bubbles connected by all the various types of interactions in our system. This is a generalized version of the ring diagram type sum.\cite{mahan2013many,fetter2012quantum,mattuck1992guide} Since the recursive relation for $\chi$ is of the form of a geometric series, we can solve for the response function outright in terms of $v$ and $\chi^{IJ}_{0~mj;ni}(1,2)$,
\begin{align}\label{eq:defchiRStoner}
\chi&^{IJ}_{mj;ni}(1,2)=\nonumber\\
&\left[ \mathbf{1} -\chi^{IM}_{0~ma;nb}(1,3)   v^{at;bs}_{ML}(3,4) \right]^{-1~IK}_{mg;nf}  \chi^{KJ}_{0~gj;fi}(4,2) .
\end{align}

In this form we gain some insight into the analytic structure of $\chi^{IJ}_{mj;ni}$. In the process of solving for $\chi^{IJ}_{mj;ni}$, we have introduced the matrix inverse of $\mathbf{1} -\mathbb{F}$ [where $\mathbb{F}=\chi^{IM}_{0~ma;nb}(1,3) v^{at;bs}_{ML}(3,4) $]. This forces the response function to be valid if and only if $\mathbf{1} -\mathbb{F}$ is nonsingular. Due to this singularity condition we can draw a few physical implications. In the static limit ($z_{2}-z_{1}\rightarrow \infty$) if $\mathbf{1} -\mathbb{F}$ becomes singular for specific periodic arrangement of $r_{1}$ and $r_{2}$, then there is an instability towards a broken symmetry phase. Physically, this means that a vanishingly small external field $\pi^{J}_{ij}(2)$ can produce an ordered state since $\chi^{IJ}_{mj;ni}>>1$,  implying the system can (wants to) lower its energy by ordering. This is a generalized Stoner criteria where charge, spin, and layer degrees of freedom mix to generate new exotic phases of matter. 

Equation~(\ref{eq:defchiRStoner}) also predicts bosonic quasiparticles as poles. Depending on the index combination of $\chi^{IJ}_{mj;ni}$, these bosons can be interlayer, intralayer, or interfacial plasmons, magnons, or coupled plasmon-magnons for noncollinear magnetic systems. The dispersion of these collective excitations  can be found by tracing energy vs momentum when the eigenvalues of $\mathbf{1} -\mathbb{F}$ equal zero.
\footnote{To find the dispersion of the collective excitations, we first use the fact that $\mathbb{F}=\sum_{i}V^{\dagger}_{i}\lambda_{i}V_{i}$, 
\begin{align}
\left[ \mathbf{1} -\mathbb{F}\right] &= \sum_{i}\left[ V^{\dagger}V -V^{\dagger}\lambda_{i}V\right]\\
&=\sum_{i}V^{\dagger}\left[ 1-\lambda_{i}\right]V,
\end{align}
where $\lambda_{i}$ is the $i$th eigenvalue of $\mathbb{F}$. Thus when $\lambda_{i}=1$, $\mathbf{1} -\mathbb{F}$ is singular and a pole is produced in $\chi^{IJ}_{jm;in}$. Therefore the dispersion of the collective mode is given by tracing energy vs momentum where $\lambda_{i}=1$. This approach is similar to those discussed in Refs.~\onlinecite{galamic2001eigenfunctions,wilson2009iterative,baldereschi1979dielectric,car1981dielectric,kaur2013spectral}.} A summary of the varous bosonic modes is given in Fig.~\ref{fig:screeningandbosons} (right panel).

Now that we have explored the structure and types of excitions harbored in $G$ and $\chi$ we can say a few words on the experimental implications. The myriad of spectroscopic probes can be classified broadly by the underlying spectral function they measure: single-particle or two-particle. The single-particle spectral function, $-\frac{1}{\pi}\text{Im}~ G$, contains all the information related to the response of a system to the removal (addition) of a single electron. This is most readily measured by angle resolved photoemission spectroscopy (ARPES).\cite{caroli1973inelastic,feibelman1974photoemission} Since the screened interaction ($W$) can be explicitly written in terms of the response function [Eq.~(\ref{eq:defchiR})] the resonant coupling between electrons and intralayer (interfacial) magnons and plasmons will appear in the ARPES spectra as waterfalls\cite{inosov2007momentum,liu2015anomalous} and kinks\cite{garcia2010through,hu2019spectroscopic} in the electronic dispersion.

In contract, the two particle spectral function is found by perturbing the ground state without changing the electron count. Here, the dynamical structure factor,\cite{sturm1993dynamic} of the physical system is measured by scattering photons,\cite{Blume1988,Grenier2014} neutrons,\cite{zaliznyak2004magnetic} and electrons\cite{sturm1993dynamic} off the sample and measuring their change in momentum and energy. Using the fluctuation dissipation theorem the dynamical structure factor is proportional to the imaginary part of the dynamical susceptibility, 
\begin{align}\label{eq:sqbare_chi}
S(\mathbf{q},\omega)=\frac{\hbar}{\pi}\frac{1}{\left( e^{-\beta\hbar\omega}-1 \right)}\text{Im}~\chi(\mathbf{q},\omega).
\end{align}
or the dielectric function, 
\begin{align}\label{eq:sqbare_dielec}
S(\mathbf{q},\omega)=\frac{\hbar q^2}{4\pi^2 e^2}\frac{1}{\left( e^{-\beta\hbar\omega} - 1\right)}\text{Im}~\varepsilon^{-1}(\mathbf{q},\omega),
\end{align}
where the charge, spin, and layer degrees of freedom have been integrated out. Therefore, intralayer, interlayer, and interfacial plasmon and magnon peaks should be present in the observed spectra. 

\begin{figure*}[t]
\includegraphics[width=0.88\textwidth]{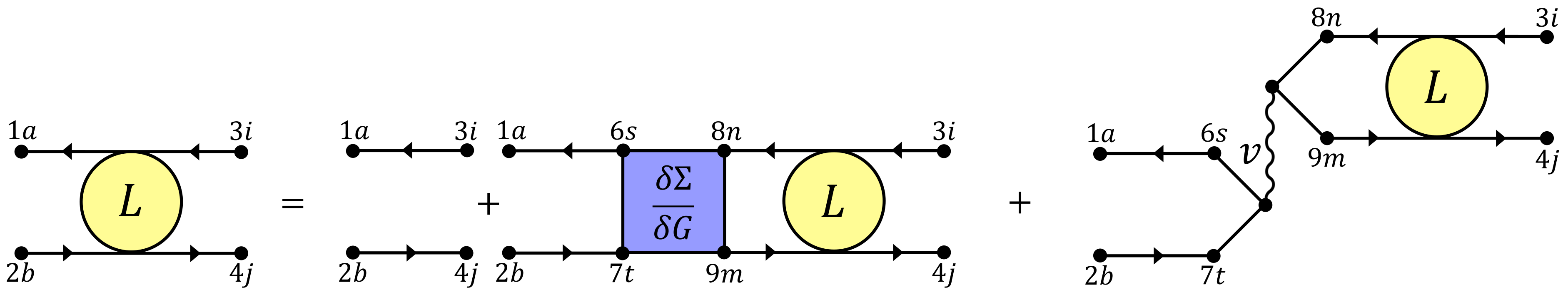}
\caption{(color online) Diagrammatic representation of the recursive relation for the two-particle propagator $L$.  The spin indices have been suppressed for clarity.}
\label{fig:2particlediagrams}
\end{figure*}

\subsection{Two-Particle Excitations and Excitons}

An important feature in the optical spectra of most semiconductors and 2D materials is the presence of electron-hole bound pairs, or {\it excitons}. The importance of excitons in 2D materials stems from their strong binding energies as a result of the highly anisotropic screening environment, leading to many novel devices and applications. \cite{mueller2018exciton,hennighausen2019widely} If we wish to characterize the various types of excitons that can form within a layered system, we need to examine the excitation spectrum of the two-particle Green's function. Our set of exact coupled equations given in Eqs.~(\ref{eq:fullsetin}) - (\ref{eq:fullsetout}), does not provide a direct means to two-particle Green's function, but rather it is recovered by judiciously combining Eqs.~(\ref{eq:vertex}) and (\ref{eq:polarization}). To directly obtain the two-particle Green's functions we first extend $\pi_{ll^{\prime}}^{I}(1)$ to a two point function, $\pi_{ll^{\prime}}^{I}(1,2)$, and consider the infinitesimal change in the single particle Green's function with respect to the two-point external field. Formally,

\begin{widetext}
\begin{subequations}
\begin{align}
\frac{\delta G_{\mu a,\nu b}(1,2)}{\delta \pi^{J}_{ij}(3,4)}
&=
-G_{\mu a,\eta s}(1,5) \frac{\delta G^{-1}_{\eta s,\xi t}(5,6)}{\delta \pi^{J}_{ij}(3,4)} G_{\xi t,\nu b}(6,2)\\
&=
-G_{\mu a,\eta s}(1,5)G_{\xi t,\nu b}(6,2) 
\left[
\frac{ \delta G^{-1}_{0~\eta s,\xi t}(5,6) }{ \delta \pi^{J}_{ij}(3,4) }
-\frac{ \delta\Sigma_{\eta s,\xi t}(5,6) }{ \delta \pi^{J}_{ij}(3,4) } 
\right] \nonumber\\
=
G_{\mu a,\eta s}(1,5)G_{\xi t,\nu b}(6,2) 
&\left[
-i\sigma_{\eta\xi}^{N} v^{ft;es}_{LN}(8,5)\sigma_{\beta\alpha}^{L} \delta(5,6)\delta(8,9) \frac{\delta G_{\beta f,\alpha e}(8,9)}{ \delta \pi^{J}_{ij}(3,4) }
+\sigma_{\eta\xi}^{J}\delta(5,3)\delta(6,4)\delta_{si}\delta_{tj}
+\frac{ \delta\Sigma_{\eta s,\xi t}(5,6) }{ \delta \pi^{J}_{ij}(3,4) } 
\right] \nonumber\\
\end{align}
\begin{align}
-i\frac{\delta G_{\mu a,\nu b}(1,2)}{\delta \pi^{J}_{ij}(3,4)}\sigma_{\mu\nu}^{I} = -iG_{\mu a,\eta i}(1,3) & \sigma_{\eta\xi}^{J}G_{\xi j,\nu b}(4,2)\sigma_{\mu\nu}^{I} \\
-i G_{\mu a,\eta s}(1,5)\sigma_{\eta\xi}^{N} & G_{\xi t,\nu b}(6,2)\sigma_{\mu\nu}^{I} 
\left[ v^{ft;es}_{LN}(8,5)\delta(5,6)\delta(8,9) 
+i\sigma_{\xi\eta}^{N}\frac{ \delta\Sigma_{\eta s,\xi t}(5,6) }{ \delta G_{\alpha e,\beta f}(8,9) }\sigma_{\alpha\beta}^{L} 
\right] (-i)\frac{\delta G_{\beta f,\alpha e}(8,9)}{ \delta \pi^{J}_{ij}(3,4) }\sigma_{\beta\alpha}^{L} \nonumber\\
\
\ \label{eq:2par}
L^{IJ}_{aj;bi}(1,2;3,4)=L^{IJ}_{0~aj;bi}(1,2;3,4)&\\
+L^{IN}_{0~at;bs}(1,2;5,6)&\left[ v^{ft;es}_{LN}(8,5)\delta(5,6)\delta(8,9) 
+i\sigma_{\xi\eta}^{N}\frac{ \delta\Sigma_{\eta s,\xi t}(5,6) }{ \delta G_{\alpha e,\beta f}(8,9) }\sigma_{\alpha\beta}^{L} 
\right]
L^{LJ}_{fj;ei}(8,9;3,4) \nonumber 
\end{align}
\end{subequations}
\end{widetext}
where $L_0$ and $L$ are the bare and dressed two-particle propagators, respectively. We should note that $L$ is a generalization of $\chi$ where the response function can be recovered by setting
\begin{align}
L^{IJ}_{aj;bi}(1,1;2,2)=\chi^{IJ}_{aj;bi}(1,2).
\end{align} 
A diagrammatic representation of the self-consistent equation for the two-particle propagator is shown in Fig.~\ref{fig:2particlediagrams}.

\begin{figure*}[t]
\includegraphics[width=0.99\textwidth]{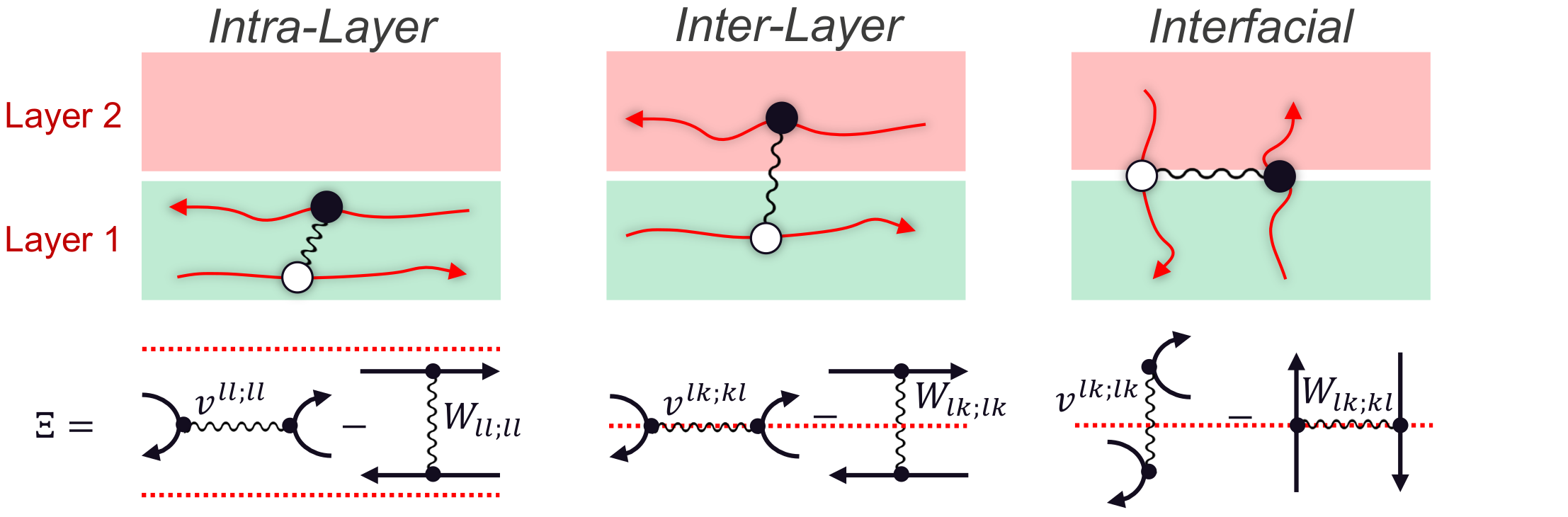}
\caption{(color online) Schematic of the various species of excitons induced by interlayer coupling, along with the diagrammatic representation of $\Xi$ for the various layer configurations. For simplicity, we have used the $GW$ approximation to evaluate $\frac{ \delta\Sigma }{\delta G}$.}
\label{fig:excitontypes}
\end{figure*}

Firstly, let us comment on the structure of Eq.~(\ref{eq:2par}). The recursion relation for $L$ takes the form of a Dyson's equation analogous to that for the dressed single-particle Green's function where the self-energy is represented by the kernel,
\begin{align}\label{eq:kernaleh}
\Xi^{NL}_{ft;es}&(5,6;8,9)= \\ &v^{ft;es}_{LN}(8,5)\delta(5,6)\delta(8,9) 
+i\sigma_{\xi\eta}^{N}\frac{ \delta\Sigma_{\eta s,\xi t}(5,6) }{ \delta G_{\alpha e,\beta f}(8,9) }\sigma_{\alpha\beta}^{L},\nonumber
\end{align}
Similar to the single-particle Green's function, we can solve for the dressed two-particle propagator $L$ in terms of $L_0$ and the `self-energy' $\Xi$ to reveal its analytic structure. Formally,
\begin{align}\label{eq:2parprop}
L^{IJ}_{aj;bi}(1,2;3,4)
=
\left[ L_{0}^{-1}-\Xi \right]^{-1~IJ}_{aj;bi}(1,2;3,4),
\end{align}
where $L_{0}^{-1}$ provides the bare two-particle excitation spectrum, and the real and imaginary part of $\Xi$ shifts the election-hole excitations and accounts for their life time, respectively. In particular, the poles of $L$, occurring when $L_{0}^{-1}-\Xi =0$, describe the pairing between and electrons and holes. 

Since we wish to characterize spin and layer dependent excitons, as seen in optical spectroscopy, we will work within the $GW$ approximation to better examine the physical content of Eq.~(\ref{eq:2parprop}) and rationalize its indexing structure. This means the kernal [Eq.~(\ref{eq:kernaleh})] reduces to the difference between the bare and screened interactions, 
\begin{align}
v^{ft;es}_{LN}(8,5)\delta(5,6)\delta(8,9) -W^{NL}_{ft;es}(6,5)\delta(5,8)\delta(6,9).
\end{align}
The screened interaction is direct and provides an attractive coupling between electrons and holes, while the bare exchange interaction is repulsive. The balance between these to opposing forces guides the creation of bound states. 

The spin structure has been analyzed by previous works\cite{rohlfing2000electron}, so we will focus on the layer degrees of freedom. Similar to the charge and magnetic response, we find three unique cases. For the case of intralayer excitons [Fig.~\ref{fig:excitontypes} (left panel)], an electron and hole propagating within the same layer $l$ interact by the screened interaction and bare exchange interaction as follows
\begin{align}
W^{ll;ll} ~~~ \text{and} ~~~ v^{ll;ll}.
\end{align}
Here, the components of $\Xi$ are very similar to the usual form employed in standard BSE calculations on bulk solids and thin films, except $\varepsilon^{-1}$ in $W$ contains the spin and charge fluctuation contributions from the surrounding layers. If the electron and hole exist on different layers [Fig.~\ref{fig:excitontypes} (center panel)], $\Xi$ takes a different form with,
\begin{align}
W^{ll;kk}~~~\text{and}~~~ v^{lk;kl}.
\end{align}
Now the exchange interaction is strictly of the layer nonconserving type, while $W$ is of the conserving type due to its direct nature. Therefore the generation of interlayer excitons is directly mediated by the competition of layer conserving and nonconserving interactions.

Finally, we predict the existence of a new type of exciton that is restricted to the interface. If electrons and holes are exchanged about the interface they can form a bound state at the boundary between the two materials [Fig.~\ref{fig:excitontypes} (right panel)]. In this case the exchange interaction is layer conserving, while the attractive screened interaction is layer nonconserving, as given by
\begin{align}
W^{lk;kl}~~~\text{and}~~~ v^{lk;lk}.
\end{align}
Interestingly, various optical spectroscopy studies have already identified interlayer excitons\cite{rivera2018interlayer,hanbicki2018double}, justifying the importance of layer nonconserving interactions in real materials.

\section{Interlayer Coupling in a Bilayer System}

As we have shown in the preceding sections interlayer coupling, through either hybridization or electron interactions, can shape and induce various plasmonic, magnonic, and excitonic excitations. In this section we will focus on a few aspects of our findings within a concrete model. Specifically, we will explore the consequence of interlayer coupling on the magnetic ordering instabilities and spin excitations in a simple Hamiltonian for a bilayer square lattice system. 

\subsection{Non-Magnetic Hamiltonian}

The Hamiltonian for a square lattice bilayer system without electron-electron interactions is explicitly written as
\begin{align}
\mathcal{H}=
\sum_{lss^{\prime}\sigma}  t^{l}_{ss^{\prime}}
c^{\dagger}_{ls\sigma}c_{ls^{\prime}\sigma}
+
\sum_{ll^{\prime}s} t^{ll^{\prime}}_{ss^{\prime}}
\left( c^{\dagger}_{ls\sigma}c_{l^{\prime}s^{\prime}\sigma} + h.c. \right),
\end{align}
where $c^{\dagger}_{ls}(c_{ls})$ create (destroy) fermions on site $s$ of layer $l$ with spin eigenvalues $\sigma=\pm$. The first term describes the hopping of electrons on each individual layer and the second term allows for hopping between the layers. Since the atomic sites within a given layer are organized over a square lattice, with full translation symmetry, we can Fourier transform the Hamiltonian of each layer. The Hamiltonian of each layer can then be expressed as  
\begin{align}
H^{l}_{\mathbf{k}}=\sum_{\sigma}\left(\sum_{\braket{ss^{\prime}}}t^{l}_{ss^{\prime}}\exp(-i\mathbf{k}\cdot \mathbf{R}_{ss^{\prime}})\right)c^{\dagger}_{l\mathbf{k}\sigma}c_{l\mathbf{k}\sigma}
\end{align}
with $\braket{ss^{\prime}}$ denoting that the sum is taken over successive rings of neighboring lattice sites surrounding site $s$, and $\mathbf{R}_{ss^{\prime}}$ is the displacement between lattice sites $s$ and $s^{\prime}$. Taking the sum out to the fourth nearest neighbor, we find the dispersion of each layer is given by 
\begin{align}\label{eq:singelbandHAM}
H^{l}_{k}=&-2t(\cos(k_{x}a)+\cos(k_{y}a))\\
&-4t^{\prime}(\cos(k_{x}a)\cos(k_{y}a))\nonumber\\
&-2t^{\prime\prime}(\cos(2k_{x}a)+\cos(2k_{y}a))\nonumber\\
&-4t^{\prime\prime\prime}(\cos(2k_{x}a)\cos(k_{y}a)+\cos(2k_{y}a)\cos(k_{x}a)),\nonumber
\end{align}
where $a$ is the lattice spacing and successive primes $(^{\prime})$ denote nearest neighbors, next-nearest neighbors, and so on. Finally, the full Hamiltonian including interlayer hybridization, or {\it bilayer splitting}, is given by
\begin{align}\label{eq:bilayerham}
H_{k\sigma}&=
\begin{bmatrix}
    H_{k\sigma}  &  t_{\perp}^k\\
    t_{\perp}^k &  H_{k\sigma} \\
\end{bmatrix},
\end{align}
where we have assumed layer one and two have the same hopping amplitudes and the interlayer hopping $t^{ll^{\prime}}_{ss^{\prime}}$ can be cast as a momentum dependent bilayer splitting $t_{\perp}^k$. 

\begin{figure}[t]
\includegraphics[width=1.0\columnwidth]{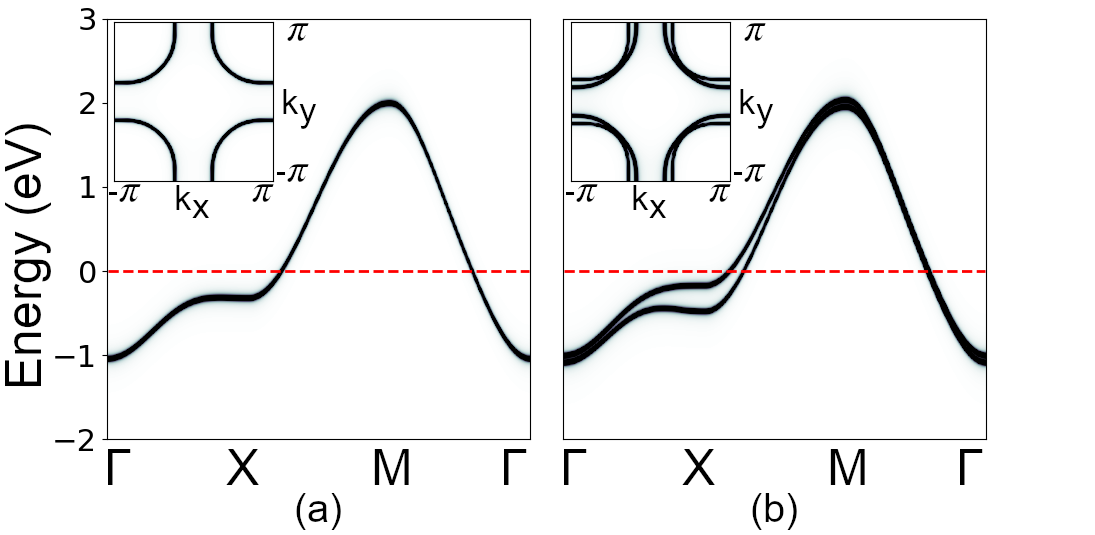}
\caption{(color online) Single particle spectral function in the absence (a) and presence (b) of bilayer splitting. The insets show the Fermi surface. }
\label{fig:NMspecfunction}
\end{figure}

Here, we will use the tight binding parametrization for the bilayer bismuth-based cuprates Bi$_2$Sr$_2$CaCu$_2$O$_{8}$ (BSCCO) as given in Ref.~\onlinecite{markiewicz2005one}, where the momentum dependent bilayer splitting is defined as 
\begin{align}
t_{\perp}^k=-t_{bi} 
\left(
\frac{ \left[ \cos(k_x a)-\cos(k_y a)\right]^2 }{4}
+
a_0
\right).
\end{align}
BSCCO, first discovered in 1988, \cite{1988_Subramarian_Science_Bi2212_discovery,1988_Tarascon_BSCCO_110K-Tc,1988_Tarascon_BSCCO_preparation_structure_properties} is one of the most studied cuprate compounds, owing to the weak van der Waals-like coupling between the rock-salt SrO-BiO$_\delta$-SrO charge reservoir layer and the two CuO$_2$-Ca-CuO$_2$ layers that facilitate cleaving for accurate surface studies with angle resolved photoemission spectroscopy~\cite{1997_Saini_PRL_ARPES_Bi2212_pseudogap_shadow_bands,2001_Ding_PRL_ARPES_Coherent_Quasiparticle_Weight,2002_Lang_Nature_Bi2212_ARPES_granular_structure,2003_Damascelli_cuprate_megareview,2015_Lanzara_ARPES_unoccupied_states,2018_Lanzara_spin-momentum-locking_Bi2212} and with scanning tunneling spectroscopy,\cite{1994_Bernardo_cuprate_gap_theoretical,2003_McElroy_Nature_Bi2212_ARPES,2007_RevModPhys_Renner_cuprate_STM,2012_Jouko_PRB_STM_VHS,2015_Mistark+Bob+Arun_Nanoscale_phase_separation_Bi2201_Ca2CuO2Cl2} 
therefore making it an interesting compound to examine the effects of interlayer coupling. The hopping parameters used  are given in Table \ref{table:hopping1}.

\begin{table}[h]
\centering
\begin{tabular}{|c|c|c|c|c|c|}
$t$ & $t^{\prime}$ & $t^{\prime\prime}$ & $t^{\prime\prime\prime}$  & $t_{bi}$ & $a_0$ \\
\hline
360 & -100 & 35 & 10 & 110 & 400\\
\end{tabular}
\caption{Tight-binding hopping parameters (in meV) for Bi$_2$Sr$_2$CaCu$_2$O$_{8}$ reproduced from Ref. \onlinecite{markiewicz2005one}.}\label{table:hopping1}
\end{table}

Figure~\ref{fig:NMspecfunction} (left panel) shows the single particle spectral function in the absence of interlayer hybridization. The band dispersion of each layer is degenerate forming a single hole-like cylinder Fermi surface centered at the corners of the Brillouin zone. For finite interlayer hybridization [Fig.~\ref{fig:NMspecfunction} (right panel)], the layer basis is reorganized into bonding and antibonding pairs, splitting the degenerate energy levels. This produces two cylindrical Fermi surfaces of slightly different doping.

\subsection{The RPA susceptibilities and magnetic ordering instabilities}\label{sec:static_susep}

To calculate the magnetic instabilities we consider the density-density response 
\begin{align}
\chi^{IJ~lj}_{0~~ki}(q,-q^{\prime},\tau)
&=\braket{T\{\hat{\sigma}^{J}_{ij}(q,\tau)\hat{\sigma}^{I}_{kl}(-q^{\prime},0)\}}
\end{align}
of the generalized density operator 
\begin{align}
\hat{\sigma}^{I}(q,\tau)_{l l^{\prime}}&=
\sum_{k}
\left(\hat{\psi}_{k+q\uparrow l}^{\dagger} \hat{\psi}_{k+q\downarrow l}^{\dagger} \right) 
\sigma^{I}  
\left( \begin{array}{c} \hat{\psi}_{k\uparrow l^{\prime}} \\ \hat{\psi}_{k\downarrow l^{\prime}} \end{array} \right),
\end{align}
where  $\tau$ is the imaginary time, $q(q^{\prime})$ is the momentum transfer, $ijkl$ index the layer, and  $I=0$ gives the charge density  and  $I={x,y,z}$ gives the spin density along each Cartesian direction. If we assume a noninteracting ground state, we can write the noninteracting susceptibilities as,
\begin{widetext}
\begin{subequations}
\begin{align}
\chi^{IJ}_0\left|^{l^{\prime}i}_{i^{\prime}l}\right.(q,i\omega_{n})&=-\sum_{k}\sum_{\substack{\alpha\beta\\ \alpha^{\prime}\beta^{\prime}}} 
\sigma^{J}_{\alpha\beta}\sigma^{I}_{\alpha^{\prime}\beta^{\prime}}
\frac{1}{\beta}\sum_{j}
G^{i^{\prime}\alpha^{\prime}i\beta}_{0}(k,i\omega_{n}+iq_{j}) 
G^{l^{\prime}\beta^{\prime}l\alpha}_{0}(k+q,iq_{j})\\
&=-\sum_{k}\sum_{\substack{\alpha\beta\\ \alpha^{\prime}\beta^{\prime}}} \sigma^{J}_{\alpha\beta}\sigma^{I}_{\alpha^{\prime}\beta^{\prime}}
\sum_{st} V_{(i\beta)s}^{k}\left(V_{(i^{\prime}\alpha^{\prime})s}^{k}\right)^{\dagger} V_{(l^{\prime}\beta^{\prime})t}^{k+q}\left(V_{(l\alpha)t}^{k+q}\right)^{\dagger}\frac{f(\varepsilon^{t}_{k+q})-f(\varepsilon^{s}_{k})}{w+\varepsilon^{t}_{k+q}-\varepsilon^{s}_{k}+i\delta}\label{eq:susep}
\end{align}
\end{subequations}
\end{widetext}
where $\beta=1/T$ and
\begin{align}\label{ex.eq:bareG}
G_{0~\alpha l, \beta k}(k,i\omega_n)=\sum_{i} \frac{V_{l\alpha,i}V^{*}_{k\beta,i}}{i\omega_n-\varepsilon_{i}}.
\end{align}
In the definition of the noninteracting Green's function [Eq.~(\ref{ex.eq:bareG})] $i\omega_n$ is the Matsubara frequency and $V_{l\alpha,i}=\braket{l\alpha|i}$ are the matrix elements connecting the layer-spin and the band spaces found by diagonalizing the Hamiltonian. The retarded susceptibility in Eq.~(\ref{eq:susep}) is found by performing the Matsubara frequency summation and by analytically continuing  $i\omega_n\rightarrow\omega+i\delta$, for $\delta\rightarrow 0^{+}$.

To calculate the charge and magnetic response functions, we consider Coulomb interactions of the electrons on the same site and between layers in an RPA framework. We distinguish between the layer conserving intralayer interaction $U$ of electrons on the same atomic site, and an interlayer interaction $V$. We also take the interfacial layer nonconserving interaction into account in two different configurations, $I$ and $I^{\prime}$, where $I$ mimics a Hund's coupling and $I^{\prime}$ describes pair hopping between the layers. Then, by including all crossed diagrams we arrive at the set of layer-dependent interactions,
\begin{subequations}
\begin{align}
U=v^{ll;ll}_{\sigma\bar{\sigma};\sigma\bar{\sigma}}=
-v^{ll;ll}_{\sigma\bar{\sigma};\bar{\sigma}\sigma}
~~&~~
I=v^{lk;kl}_{\sigma\sigma;\sigma\sigma}=
-v^{lk;lk}_{\sigma\sigma;\sigma\sigma}
\\
V=v^{lk;lk}_{\sigma\sigma;\sigma\sigma} =
-v^{lk;kl}_{\sigma\sigma;\sigma\sigma}     
~~&~~
I=v^{lk;kl}_{\sigma\bar{\sigma};\sigma\bar{\sigma}}=
-v^{lk;lk}_{\sigma\bar{\sigma};\bar{\sigma}\sigma}
\\ 
V=v^{lk;lk}_{\sigma\bar{\sigma};\sigma\bar{\sigma}}=
-v^{lk;kl}_{\sigma\bar{\sigma};\bar{\sigma}\sigma} 
~~&~~
I^{\prime}=v^{l l;k k}_{\sigma\bar{\sigma};\sigma\bar{\sigma}}=
-v^{ll;kk}_{\sigma\bar{\sigma};\bar{\sigma}\sigma},
\end{align}
\end{subequations}
Finally, in Table \ref{example:table:pauli_interactions} we expand the interactions in the Pauli basis. Since the interactions do not contain any spin flips, only the $v^{00}$, $v^{xx}$, $v^{yy}$, and $v^{zz}$ terms are nonzero.  Consequently, the interactions are rotationally invariant for each layer dependent configuration.
\begin{table}[h]\label{tab:intrpa}
\centering
{\renewcommand{\arraystretch}{1.8}
\begin{tabular}{c|c|c|c|c|}
&$v^{00}$&$v^{xx}$&$v^{yy}$&$v^{zz}$\\\hline\hline
$~v^{ll;ll}~$& $\frac{U}{2}$ 				& $-\frac{U}{2}$ 			& $-\frac{U}{2}$ 			& $-\frac{U}{2}$ \\\hline
$~v^{lk;lk}~$& $V-\frac{I}{2}$	& $-\frac{I}{2}$ 			& $-\frac{I}{2}$ 			& $-\frac{I}{2}$\\\hline
$~v^{kl;lk}~$& $I-\frac{V}{2}$	& $-\frac{V}{2}$ 	& $-\frac{V}{2}$ 	& $-\frac{V}{2}$\\\hline
$~v^{kk;ll}~$& $\frac{I^{\prime}}{2}$		& $-\frac{I^{\prime}}{2}$ 	& $-\frac{I^{\prime}}{2}$ 	& $-\frac{I^{\prime}}{2}$ \\\hline
\end{tabular}
}
\caption{Layer components of the interaction in the Pauli basis $(l\neq k)$.}\label{example:table:pauli_interactions}
\end{table}

Using the noninteracting single-particle propagator [Eq.~(\ref{ex.eq:bareG})] in the polarization $\chi_{0}$ along with taking the bare vertex [Eq.~(\ref{eq:bareVertex})] in the layer-dependent electromagnetic response function $\chi^{IJ}_{l^{\prime}k;k^{\prime}l}$ we recover the generalized RPA susceptibilities,
\begin{align}\label{eq:rpasusp}
&\chi^{IJ}_{l^{\prime}k;k^{\prime}l}(q,\omega)=\\[0.75em]
&\chi^{IJ}_{0~l^{\prime}k;k^{\prime}l}(q,\omega)
+
\chi^{IK}_{0~l^{\prime}m;k^{\prime}n}(q,\omega)
v^{mm^{\prime};nn^{\prime}}_{KL}
\chi^{LJ}_{m^{\prime}k;n^{\prime}l}(q,\omega),\nonumber
\end{align}
where repeated indices are summed over. For a single-band susceptibility the inclusion of interactions within the RPA approach enhances existing features in the noninteracting susceptibility as the Stoner denominator $1-U\chi(q,\omega)$ approaches zero. In the case of a multilayer susceptibility, much like the multiorbital case,\cite{graser2009near} it is not obvious how the different structures in the spin and in the charge susceptibility are changed by the varying $U$, $V$, $I$, and $I^{\prime}$. To present a simplified and transparent discussion, we varied each parameter while tracking various spin correlation functions.

\begin{figure}[h!]
\includegraphics[width=1.0\columnwidth]{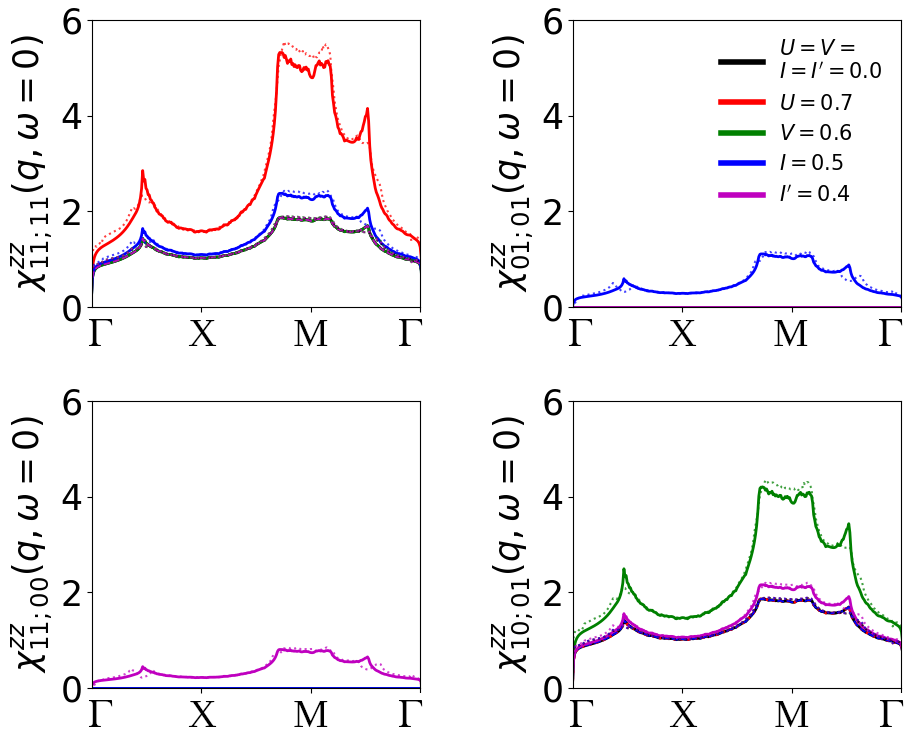}
\caption{(color online) The generalized RPA spin susceptibilities calculated with (dotted lines) and without interlayer hybridization (solid lines).  }
\label{fig:spincorrelations}
\end{figure}

Figure~\ref{fig:spincorrelations} shows the RPA spin correlations along the high-symmetry line in the square Brillouin zone for intralayer $\braket{S_{11}S_{11}}$, interlayer $\braket{S_{00}S_{11}}$, and interfacial $\braket{S_{01}S_{01}}$ and $\braket{S_{01}S_{10}}$ spin configurations with (dotted lines) and without (solid lines) interlayer hybridization. For $U=0.7$ eV there is a dramatic enhancement in the spin susceptibilities near $M$ in the intralayer channel. This enhancement signals an instability toward $(\pi,\pi)$ AFM order, which is in agreement with other RPA studies of cuprates\cite{markiewicz2017entropic} and the experimentally observed AFM order in the BSCCO parent compound.\cite{kastner1998magnetic} Upon introducing $V$,  there is an increase in the spin fluctuations at $(\pi,\pi)$  in the $\braket{S_{00}S_{11}}$ channel, similar to the effect of $U$. Physically, this interlayer interaction gives rise to the various AFM orderings along the $c$ axis, e.g., G- and C-type AFM orders. Finally, for finite $I$ and $I^{\prime}$, spin correlations appear in the $\braket{S_{01}S_{01}}$ and $\braket{S_{01}S_{10}}$ sectors. This suggests the existence of instabilities towards interfacial magnetic ordering in the BSCCO bilayer system.  Following the dotted lines, we find that a finite interlayer hybridization tends to round-out nonanalytic cusps and plateaus near $M$, eliminating competition between various AFM orders, as expected for systems in more than two dimensions.\cite{markiewicz2017entropic,markiewicz2004mode}

\subsection{Antiferromagnetic Hamiltonian and Induced Magnetic Order}\label{sec:dynamic_susep}

Proximity effects play a significant role in designing new functional heterostructures with strategically induced phases such as superconductivity,\cite{di2017p,zareapour2012proximity} spin-orbit coupling effects,\cite{wang2015strong,zhou2019spin} and magnetism.\cite{lee2016direct,zollner2019proximity,hou2019magnetizing} Specifically, in the layered cuprate high-temperature superconductors extensive NMR studies on multilayer cuprates have observed an inhomogeneous hole doping of the various CuO$_2$ layers, resulting in the coexistence of nearly pristine and optimally doped CuO$_2$ planes. \cite{mukuda2006uniform,mukuda2011high} In the case of single-layered cuprates the relative hole doping between layers can be manipulated through the so-called $\delta$-doping scheme.\cite{suter2018superconductivity} This presents a natural platform to explore the role hybridization and layer-dependent interactions independently play in these proximity effects. 

To explore this, we introduce a $Q=(\pi,\pi)$ AFM order into one of the layers in our Hamiltonian for bilayer BSCCO [Eq.~\ref{eq:bilayerham}] mimicking the inhomogeneous hole doping observed experimentally. After taking the Umklapp processes into account and factoring the electron-electron interactions through an auxiliary field, we arrive at the Hamiltonian in terms of the self-consistent field $m$ and occupation $n_{\sigma}$,
\begin{align}\label{eq:scham}
H_{k\sigma}&=\begin{bmatrix}
    H_{k\sigma}                   &  sign(\bar{\sigma})\Delta           &  t_{\perp}^k  & 0\\
    sign(\bar{\sigma})\Delta      &  H_{k+Q\sigma}                	   &  0          & t_{\perp}^{k+Q}\\
    t_{\perp}^{k}                     & 0                                   & H_{k\sigma} & 0\\
    0                             & t_{\perp}^{k+Q}                           & 0           &  H_{k+Q\sigma} \\
\end{bmatrix}
\end{align}
where our wave functions take the Nambu form $\Psi^\dagger=\left( c^{\dagger}_{1k\sigma}~,~ c^{\dagger}_{1k+Q\sigma} ~,~  c^{\dagger}_{2k\sigma}~,~ c^{\dagger}_{2k+Q\sigma} \right)$, $\Delta$ is defined as $\frac{U}{2}\left(m+m^{\dagger}\right)=URe\left(m\right)$, and the constant shift $Un_{\bar{\sigma}}	$ is added to the chemical potential. See Appendix~\ref{A:meanfield} for a detailed derivation of the mean-field Hamiltonian. 

\begin{figure}[h!]
\includegraphics[width=0.85\columnwidth]{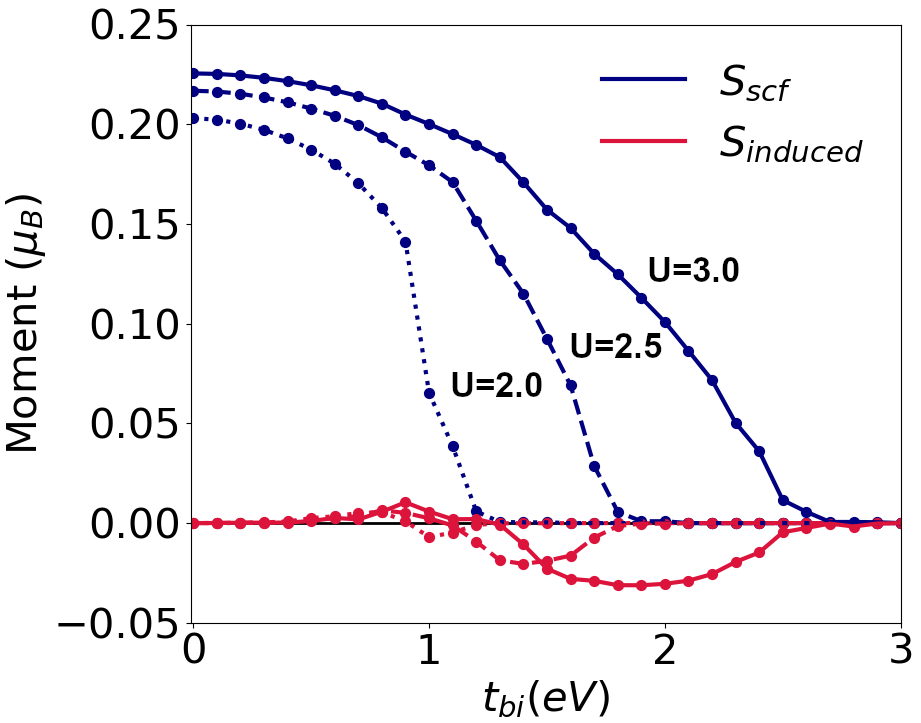}
\caption{(color online) The self-consistent magnetic moment $S_{scf}$ and the induced magnetic moment $S_{induced}$ as a function of interlayer hybridization for three different onsite correlation strengths $U$.}
\label{fig:magmom_interlayer}
\end{figure}

Figure~\ref{fig:magmom_interlayer} shows the self-consistent spin magnetic moment $S_{scf}$ on layer 1 and the induced magnetic moment $S_{induced}$ in layer 2 as a function of bilayer splitting for three different onsite correlation strengths $U$. For a $U$ of $2.0$ eV, $S_{scf}$ has a maximum of 0.203 $\mu_B$ with no bilayer splitting. For finite $t_{bi}$, three distinct regions are observed. (I) For $0~ \leq t_{bi}\leq 0.9$ eV a positive $S_{induced}$ is produced, reaching a maximum of $0.006$ $\mu_B$. (II) When $0.9 \leq t_{bi}\leq 1.3$ eV, $S_{induced}$ is negative with a minimum of $-0.008$ $\mu_B$. (III) For $1.3 \leq t_{bi}$ eV both $S_{scf}$ and $S_{induced}$ are quenched.  $S_{scf}$ decreases for increasing values of $t_{bi}$, with a visible kink in the line shape concomitant with the change in sign of $S_{induced}$. For larger onsite potentials, the region  and moment of negative $S_{induced}$ is increased and enhanced, respectably.

\begin{figure}[t]
\includegraphics[width=1.0\columnwidth]{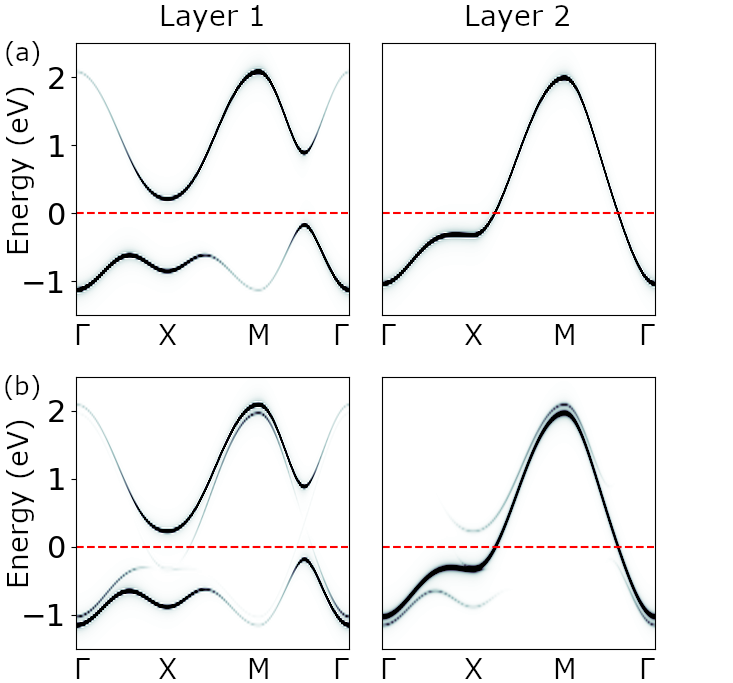}
\caption{(color online) The single-particle spectral function of a bilayer AFM-metallic system with and without interlayer hybridization. (Top) Shows the spectral weight for layer 1 with $(\pi,\pi)$ AFM order and an uncorrelated metallic layer 2. (bottom) Shows the effect of interlayer hybridization on the spectra of each layer.}
\label{fig:bands_interlayer}
\end{figure}

Physically, the increase in $t_{bi}$ can be facilitated by uniaxial compressive strain
where the two CuO$_2$ layers are brought into closer proximity, allowing greater wave function overlap. The change in sign of $S_{induced}$ suggests a change from C-type to G-type AFM order purely due to hybridization. A similar type of  behavior is observed in the bilayer CrI$_3$ where different layer stacking configurations induce AFM or FM coupling between the layers.\cite{sivadas2018stacking} Furthermore, the delicate interlayer hopping between IrO$_6$ planes in Sr$_2$IrO$_4$ can be disrupted by an external laser pulse, changing the magnetic symmetry of the system.\cite{di2016magnetic,takayama2016model,zhao2016evidence,lane2020iridate}

\begin{figure*}[t]
\includegraphics[width=1.0\textwidth]{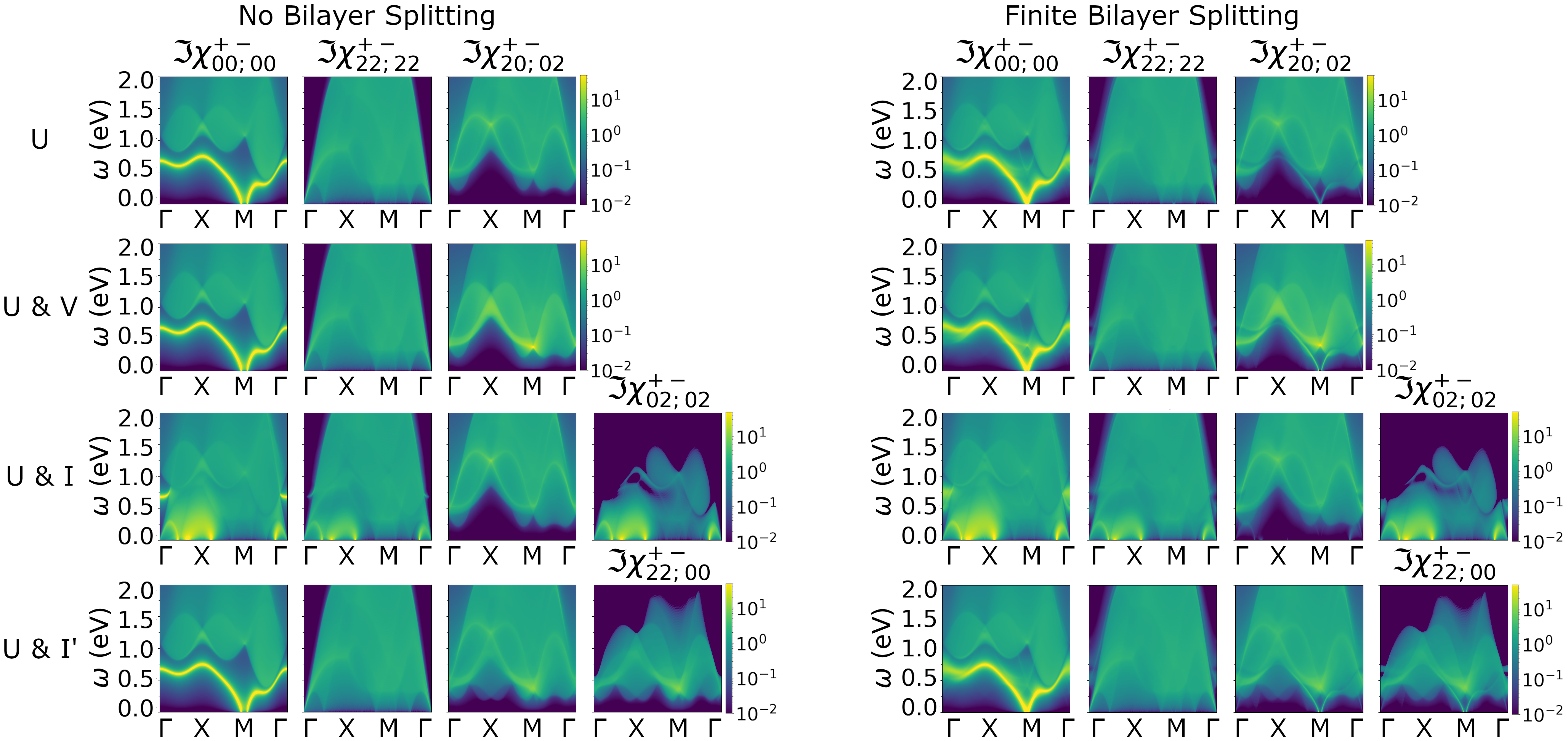}
\caption{(color online) The relevant nonzero tensor components of $\Im\chi^{+-}(q,\omega)$ for various layer-dependent interactions in the absence (left panel) and presence (right panel) of bilayer splitting. The layer-dependent interactions used in each row are noted on the left.} 
\label{fig:chi_pm}
\end{figure*}

Figure~\ref{fig:bands_interlayer} (top) shows the single-particle spectral function without bilayer splitting. Layer 1 exhibits a 1 eV AFM band gap at the $X$ point and along the $M-\Gamma$ direction in the square Brillouin zone. Since correlations were turned off in layer 2, it is a metal.  The bottom panels of Fig.~\ref{fig:bands_interlayer} display the effect of interlayer hybridization on the single particle states. Firstly, the spectra of both layers is present in the projected spectral weight of each layer. This is produced by $t_{bi}$ forming bonding (antibonding) pairs between various layer quantum numbers. Moreover, the wide AFM gap of layer 1 is clearly seen, along with a very slight induced gap produced in the originally metallic band of layer 2.  

\subsection{Spin Waves in the Presence of Interlayer Coupling}

To mark the effect of the various layer dependent interactions and bilayer splitting on the spin wave dispersion in layer 1 and metallic character of layer 2, we calculate the imaginary part of the transverse spin susceptibility $\Im\chi^{+-}(q,\omega)$ in the random phase approximation [Eq.~(\ref{eq:rpasusp})]. {\color{red} } The results are presented in Figure.~\ref{fig:chi_pm} and are organized as follows. The right and left panels show $\Im\chi^{+-}(q,\omega)$ with and without bilayer splitting, respectively. The rows in each panel present the data for the various layer-dependent interaction combinations used in the RPA. The specific interactions used are noted on the left. The values of $U$, $V$, $I$, and $I^{\prime}$, employed are $1.5$ eV, $1.0$ eV, $1.0$ eV, and $1.0$ eV, respectively. For brevity, only the relevant nonzero tensor components are given.

 Figure~\ref{fig:chi_pm} (left panel, row one) shows the nonzero components of $\Im\chi^{+-}(q,\omega)$ along high symmetry lines in the Brillouin zone for just an onsite potential $U$. In layer 1 (channel $00;00$), a clear gapless spin wave dispersion is observed, with its energy minimum at $(\pi,\pi-\delta)$. The spin excitation is clear throughout the Brillouin zone, never entering the continuum and damping out. In contrast, layer 2 (channel $22;22$) exhibits a gapless particle-hole continuum, consistent with its metallic band structure. Furthermore, the interfacial channel $20;02$ is nonzero exhibiting a faint gapped spin excitation band at $0.35$ eV on top of the particle-hole continuum. When interlayer and interfacial interactions $V$ (row two) and $I^{\prime}$ (row four) are introduced, the spectra is relatively unchanged except for an enhancement in the gapped interfacial spin mode in channel $20;02$. Lastly, $I^{\prime}$ generates a new nonzero interfacial matrix element (channel $22;00$) with a similar structure to that of channel $20;02$.
 
Interestingly, a finite layer nonconserving interaction $I$  (row three) dramatically damps the magnon dispersion in layer 1 by mixing in the metallic particle-hole continuum of layer 2. Moreover, the zero of the dispersion is shifted to surrounding $\Gamma$ and $X$. Since $I$ mixes the excitation spectrum of layer 1 and 2, a magnon dispersion is now induced in layer 2, similar to layer 1. Additionally, a new interfacial nonzero channel $02;02$  is found, displaying characteristic features of layer 1 and 2.

\begin{figure}[ht]
\includegraphics[width=1.0\columnwidth]{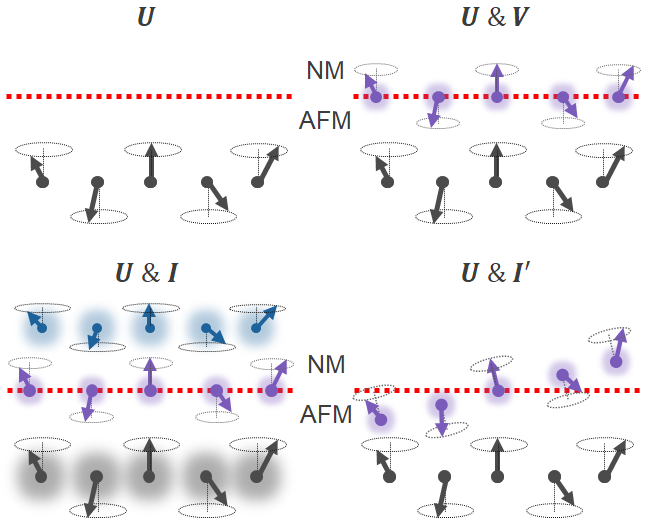}
\caption{(color online) Schematic of the various spin waves induced by different combinations of intralayer, interlayer, and interfacial interactions. The red dashed line denotes the boundary between the nonmagnetic (NM) and antiferromagnetic (AFM) layers. The cloud surrounding the magnetic moments indicates if the spin wave is damped.  }
\label{fig:chischmatic}
\end{figure}

If a finite bilayer splitting is included (right panel), the spectrum of $\Im\chi^{+-}(q,\omega)$ is very similar to that of the isolated case, except for a few key aspects. The magnon dispersion in layer 1 [seen in row one, two, and four] now has its minimum at the $M$ point in the Brillouin zone. Moreover, along $X-\Gamma$ an avoided crossing appears and the magnon mode becomes incoherent near $\Gamma$, due to the admixture of metallic features from layer 2. In the interfacial channel $20;02$, the spin wave band from layer one is clearly seen extending into the continuum. Figure~\ref{fig:chischmatic} shows a schematic summary of the various spin waves induced by the different combinations layer dependent interactions.

Lastly, through the dynamical structure factor $S(\mathbf{q},\omega)$ [Eq.~(\ref{eq:sqbare_chi})] many of the key features of layer dependent interactions appearing in the magnetic instabilities and modifications to the spin wave spectrum shown in Figs.~\ref{fig:spincorrelations} and \ref{fig:chi_pm} are directly accessible to neutron and x-ray scattering. Therefore, the prediction of layer nonconserving induced interfacial magnetic order and propagating spin waves can be readily confirmed.

In summary, layer-dependent interactions are able to modify magnetic ordering tendencies and magnon dispersions, and induce collective modes in neighboring layers all without interlayer hybridization. This illustrates the key role these interactions play in designing and manipulating various charge and magnetic phases and excitations in 2D atomically-thin film heterostructures and layered correlated compounds, such as the perovskite transition-metal oxides. 

\section{Concluding Remarks}
We have derived a generalization of Hedin's equations for a layered system with arbitrarily strong interlayer coupling. Our approach was made sufficiently general to accommodate nonlocal interactions and nonequilibrium quantum phases through the Keldysh and Schwinger techniques. We have thus opened a pathway for examining the interplay of charge, spin, orbital and layer degrees of freedom in layered heterostructures and their phase diagrams including relativistic magnetic interactions, along with the evolution of electronic spectra with pressure and doping.

\begin{acknowledgments}
The author would like to thank Dr. Zachariah Hennighausen and Dr. Jian-Xin Zhu for many fruitful discussions. This work was carried out under the auspices of the U.S. Department of Energy (DOE) National Nuclear Security Administration under Contract No. 89233218CNA000001. It was supported by the LANL LDRD Program, and in part by the Center for Integrated Nanotechnologies, a DOE BES user facility, in partnership with the LANL Institutional Computing Program for computational resources.
\end{acknowledgments}

\appendix

\section{Mean-Field Interactions and AFM Order}\label{A:meanfield}

In order to include staggered AFM order on the atomic sites, we include an onsite Hubbard interaction term to the Hamiltonian of Eq.~(\ref{eq:bilayerham}). Specifically, the double-occupancy energy penalty $U$ is placed on the single effective band crossing the Fermi level. The Hubbard interaction can be written in momentum space as
\begin{align}
\frac{U}{2}\sum_{\sigma}\sum_{kk^{\prime}Q} c^{\dagger}_{k\sigma} c_{k\sigma} c^{\dagger}_{k^{\prime}\bar{\sigma}} c_{k^{\prime}\bar{\sigma}}+c^{\dagger}_{k+Q\sigma} c_{k\sigma} c^{\dagger}_{k^{\prime}\bar{\sigma}} c_{k^{\prime}+Q\bar{\sigma}},
\end{align}
where $\bar{\sigma}$ denotes $-\sigma$. Due to momentum conservation, the interaction depends on both the crystal momentum, $k(k^{\prime})$, of the electrons and the momentum transferred, $Q$, during the interaction. The momentum transfer gives rise to Umklapp processes where electrons can scatter to neighboring Brillouin zones, which are the key for describing various density-wave instabilities. Here we take $Q=(\pi,\pi)$ following the experimentally observed AFM order. Thus, the full single-band Hamiltonian is  
 \begin{align}\label{Eq:Hamint}
\mathcal{H}&=
\sum_{\sigma}\sum_{k} \left( H_{k\sigma} c^{\dagger}_{k\sigma}c_{k\sigma} + H_{k+Q\sigma} c^{\dagger}_{k+Q\sigma}c_{k+Q\sigma} \right)\nonumber\\
&-\mu\sum_{\sigma}\sum_{k}\left(\hat{n}_{k\sigma}+\hat{n}_{k+Q\sigma} \right)\nonumber\\
&+\frac{U}{2}\sum_{\sigma}\sum_{kk^{\prime}} c^{\dagger}_{k\sigma} c_{k\sigma} c^{\dagger}_{k^{\prime}\bar{\sigma}} c_{k^{\prime}\bar{\sigma}}+c^{\dagger}_{k+Q\sigma} c_{k\sigma} c^{\dagger}_{k^{\prime}\bar{\sigma}} c_{k^{\prime}+Q\bar{\sigma}}.
\end{align}
where $H_{k}$ is written in terms of $Q$ explicitly by restricting $k(k^{\prime})$ to the smaller AFM Brillouin zone. We now rewrite the interaction in terms of the mean field and expand the number operator in terms of fluctuations away from the mean electron count per state, $\braket{n_{k\sigma}}$:
 \begin{align}
n_{k\sigma} &= \braket{n_{k\sigma}}+\left( n_{k\sigma}-\braket{n_{k\sigma}} \right)\\
&= \braket{n_{k\sigma}}+\delta_{\sigma},\nonumber
\end{align}
where $\delta_{\sigma}$ is the fluctuation away from $\braket{n_{k\sigma}}$. We substitute into the interaction of Eq.~(\ref{Eq:Hamint}) assuming fluctuations are small, $\delta_{\sigma}\delta_{\bar{\sigma}}\approx 0$, giving
\begin{align}\label{eq:meanmat}
\frac{U}{2}\sum_{\sigma}\sum_{kk^{\prime}} \braket{c^{\dagger}_{k\sigma} c_{k\sigma}}  c^{\dagger}_{k^{\prime}\bar{\sigma}} c_{k^{\prime}\bar{\sigma}}+\braket{c^{\dagger}_{k^{\prime}\bar{\sigma}} c_{k^{\prime}\bar{\sigma}}} c^{\dagger}_{k\sigma} c_{k\sigma}\\
+\braket{c^{\dagger}_{k+Q\sigma} c_{k\sigma}} c^{\dagger}_{k^{\prime}\bar{\sigma}} c_{k^{\prime}+Q\bar{\sigma}}\nonumber+\braket{c^{\dagger}_{k^{\prime}\bar{\sigma}} c_{k^{\prime}+Q\bar{\sigma}}} c^{\dagger}_{k+Q\sigma} c_{k\sigma}\nonumber.
\end{align}
In order to treat the various matrix elements in Eq.~(\ref{eq:meanmat}), we consider the average charge and spin densities as a function of momentum transfer $q$,
\begin{subequations}
\begin{align}
\braket{\rho(q)}&=\sum_{k}\braket{\left(c_{k+q\uparrow}^{\dagger} c_{k+q\downarrow}^{\dagger} \right) \mathbb{I}  \left( \begin{array}{c} c_{k\uparrow} \\ c_{k\downarrow} \end{array} \right)}\\
&=\sum_{k}\braket{c_{k+q\uparrow}^{\dagger}  c_{k\uparrow}}+\braket{c_{k+q\downarrow}^{\dagger}  c_{k\downarrow}}\nonumber\\
&=N_{e}\delta_{q,0}\nonumber\\
\braket{S^{z}(q)}&=\frac{1}{2}\sum_{k}\braket{\left(c_{k+q\uparrow}^{\dagger} c_{k+q\downarrow}^{\dagger} \right) \sigma^{z}  \left( \begin{array}{c} c_{k\uparrow} \\ c_{k\downarrow} \end{array} \right)}\\
&=\frac{1}{2}\sum_{k} \braket{c_{k+q\uparrow}^{\dagger}c_{k\uparrow}}-\braket{c_{k+q\downarrow}^{\dagger}c_{k\downarrow}}.  \nonumber
\end{align}
\end{subequations}
Therefore, for $q=Q=(\pi,\pi)$,
\begin{align}
\braket{\rho(Q)}&=\sum_{k}\braket{c_{k+Q\uparrow}^{\dagger}  c_{k\uparrow}}+\braket{c_{k+Q\downarrow}^{\dagger}  c_{k\downarrow}}\\
&=0\nonumber
\end{align}
which implies,
\begin{align}\label{eq:homoresult}
\braket{c_{k+Q\uparrow}^{\dagger}  c_{k\uparrow}}=-\braket{c_{k+Q\downarrow}^{\dagger}  c_{k\downarrow}}.
\end{align}
Also, by hermiticity we have the equivalence, 
\begin{align}
\braket{c_{k+Q\sigma}^{\dagger}  c_{k\sigma}}^{\dagger}=\braket{c_{k\sigma}^{\dagger}  c_{k+Q\sigma} }.
\end{align}
Using the relation in Eq.~(\ref{eq:homoresult}) we find $\braket{S^{z}(Q)}$,
\begin{align}
\braket{S^{z}(Q)}&=\frac{1}{2}\sum_{k} \braket{c_{k+Q\uparrow}^{\dagger}c_{k\uparrow}}-\braket{c_{k+Q\downarrow}^{\dagger}c_{k\downarrow}}\\
&=\sum_{k} \braket{c_{k+Q\uparrow}^{\dagger}c_{k\uparrow}}.\nonumber
\end{align}
The preceding relations allow us to cast staggered magnetization and electron density as,
\begin{subequations}
\begin{align}
m&=\sum_{k} \braket{c_{k+Q\uparrow}^{\dagger}c_{k\uparrow}}=-\sum_{k} \braket{c_{k+Q\downarrow}^{\dagger}c_{k\downarrow}},\\
n_{\sigma}&=\sum_{k} \braket{c_{k\sigma}^{\dagger}c_{k\sigma}}.
\end{align}
\end{subequations}
Inserting these definitions and simplifying we arrive at the Hamiltonian in terms of the self-consistent field $m$ and occupation $n_{\sigma}$,
\begin{align}\label{eq:scham}
H_{k\sigma}&=\begin{bmatrix}
    H_{k\sigma}+Un_{\bar{\sigma}}&sign(\bar{\sigma})\Delta\\
    sign(\bar{\sigma})\Delta&  H_{k+Q\sigma}+Un_{\bar{\sigma}}  \\
\end{bmatrix},
\end{align}
where our wave functions take the Nambu form $\Psi=\left( c^{\dagger}_{k\sigma}~,~ c^{\dagger}_{k+Q\sigma} \right)$ and $\Delta$ is defined as $\frac{U}{2}\left(m+m^{\dagger}\right)=URe\left(m\right)$.

To self consist $m$ and $n$, their expectation value can be written in terms of the diagonalized system. Let the quasiparticle creation $(\gamma_{ki}^{\dagger})$ and annihilation $(\gamma_{ki})$, operators in the diagonalized system be defined as
\begin{align}
c_{k\sigma}&=\sum_{i}V^{k}_{\sigma,i}\gamma_{ki} ~~\text{and}~~ c_{k\sigma}^{\dagger}=\sum_{i}\gamma^{\dagger}_{ki}(V^{k}_{\sigma,i})^{\dagger},
\end{align}
where $i$ indexes the bands. Therefore $m$ and $n$ are given by
\begin{subequations}
\begin{align}
n_{\sigma}&=\sum_{i}\sum_{k} \left( (V^{k}_{\sigma i})^{\dagger}    V^{k}_{ \sigma i} + (V^{k+Q}_{\sigma i})^{\dagger}    V^{k+Q}_{\sigma i}  \right) 
f(\epsilon_{k\sigma i}),\\
m&=\sum_{i}\sum_{k} \left( (V^{k+Q}_{\sigma i})^{\dagger}    V^{k}_{ \sigma i} + (V^{k+Q}_{\sigma i})^{\dagger}    V^{k+Q}_{\sigma i}  \right) 
f(\epsilon_{k\sigma i})
\end{align}
\end{subequations}
for $k$ in the AFM Brillouin zone and $f$ being the Fermi function. The self-consistently obtained values of the expectation value of $m$ and $n_{\sigma}$ are calculated within a tolerance of $10^{-5}$ at a temperature of $0.001$ K.

\bibliography{InterlayerCoupling_Refs}

\end{document}